\documentclass[preprint]{elsarticle}

\usepackage{lineno,hyperref}
\usepackage{graphicx}
\usepackage{amsmath,,amsfonts,amssymb}
\usepackage{booktabs}
\usepackage{tabularx}
\usepackage{multirow}
\usepackage{makecell}
\usepackage{caption}
\usepackage{listings}
\usepackage{graphicx}
\modulolinenumbers[5]

\journal{Neurocomputing}

\renewcommand\vec[1]{\overrightarrow{#1}}
\newcommand\cev[1]{\overleftarrow{#1}}

\AtBeginDocument{\renewcommand{\ref}[1]{\autoref{#1}}}
\newcolumntype{R}[1]{>{\raggedleft\hsize=#1\hsize\arraybackslash}X}
\newcolumntype{L}[1]{>{\raggedright\hsize=#1\hsize\arraybackslash}X}
\newcolumntype{C}[1]{>{\centering\hsize=#1\hsize\arraybackslash}X}

\begin{document}

\begin{frontmatter}

\title{Multimodal deep learning for short-term stock volatility prediction}

%% Group authors per affiliation:
%\author{Elsevier\fnref{myfootnote}}
%\address{Radarweg 29, Amsterdam}
%\fntext[myfootnote]{Since 1880.}
%
%%% or include affiliations in footnotes:
%\author[mymainaddress,mysecondaryaddres\includegraphics{universe}s]{Elsevier Inc}
%\ead[url]{www.elsevier.com}
%
%\author[mysecondaryaddress]{Global Customer Service\corref{mycorrespondingauthor}}
%\cortext[mycorrespondingauthor]{Corresponding author}
%\ead{support@elsevier.com}
%
%\address[mymainaddress]{1600 John F Kennedy Boulevard, Philadelphia}
%\address[mysecondaryaddress]{360 Park Avenue South, New York}

\author[add1]{Marcelo Sardelich\corref{cor1}}
\ead{marcelo.sardelich@york.ac.uk}
\author[add1]{Suresh Manandhar}
\ead{suresh@cs.york.ac.uk}
%\author[add1]{Dimitar Kazakov}
%\ead{dimitar.kazakov@york.ac.uk}

\cortext[cor1]{Corresponding author}
\address[add1]{Department of Computer Science Deramore Lane, University of York, Heslington, York, YO10 5GH, UK}

\begin{abstract}
Stock market volatility forecasting is a task relevant to assessing market risk. We investigate the interaction between news and prices for the one-day-ahead volatility prediction using state-of-the-art deep learning approaches. The proposed models are trained either end-to-end or using sentence encoders transfered from other tasks. We evaluate a broad range of stock market sectors, namely Consumer Staples, Energy, Utilities, Heathcare, and Financials. Our experimental results show that adding news improves the volatility forecasting as compared to the mainstream models that rely only on price data. In particular, our model outperforms the widely-recognized GARCH(1,1) model for all sectors in terms of coefficient of determination $R^2$, $MSE$ and $MAE$, achieving the best performance when training from both news and price data.
\end{abstract}


\end{frontmatter}

%\linenumbers

\section{Introduction}
\label{sec:introduction}
Natural Language Processing (NLP) has increasingly attracted the attention of the financial community. This trend can be explained by at least three major factors. The first factor refers to the business perspective. It is the economics of gaining competitive advantage using alternative sources of data and going beyond historical stock prices, thus, trading by analyzing market news automatically. The second factor is the major advancements in the technologies to collect, store, and query massive amounts of user-generated data almost in real-time. The third factor refers to the progress made by the NLP community in understanding unstructured text. %Initially, the NLP field relied on hand-crafted ways of representing textual features, which depend on counting word terms and weighting these terms for each specific application. This approach could not take into account the order of words in a sentence. In the past decade, the field evolved to a data-driven approach that uses deep neural networks \cite{Bengio2013}, where, usually, the building blocks are dense word embeddings able to capture the word context \cite{MikolovNIPS2013}.%

Over the last decades the number of studies using NLP for financial forecasting has experienced exponential growth. According to \cite{Xing2018}, until 2008, less than five research articles were published per year mentioning both ``stock market'' and ``text mining'' or ``sentiment analysis'' keywords. In 2012, this number increased to slightly more than ten articles per year. The last numbers available for 2016 indicates this has increased to sixty articles per year.

The ability to mechanically harvest the sentiment from texts using NLP has shed light on conflicting theories of financial economics. Historically, there has been two differing views on whether disagreement among market participants induces more trades. The ``non-trade theorem'' \cite{Milgrom1982} states that assuming all market participants have common knowledge about a market event, the level of disagreement among the participants does not increase the number of trades but only leads to a revision of the market quotes. In contrast, the theoretically framework proposed in \cite{Harris1993} advocates that disagreement among market participants increases trading volume. Using textual data from Yahoo and \href{https://ragingbull.com/}{RagingBull.com} message boards to measure the dispersion of opinions (positive or negative) among traders, it was shown in \cite{Antweiler2004} that disagreement among users' messages helps to predict subsequent trading volume and volatility. Similar relation between disagreement and increased trading volume was found in \cite{Sprenger2014} using Twitter posts. %The dispersion index proposed in the work is given by  $B^{t} = (M_{t}^{P} - M_{t}^{N}) / (M_{t}^{P} + M_{t}^{N})$, where $M_{t}$ is the weighted sum of positive ($P$) and negative ($N$) news in a given day $t$. A value of $B^{t}$ close to zero means a balanced market opinion among the message board users while the extreme values +1 (-1) a market dominated by optimism (pessimism). To perform the sentiment analysis of each post a small set of messages was manually annotated and trained using a Naive Bayes classifier.

Additionally, textual analysis is adding to the theories of medium-term/long-term momentum/reversal in stock markets \cite{Vayanos2013}. The unified Hong and Stein model\footnote{The \textit{gradual information diffusion} model of Hong and Stein considers two types of economic agents, namely ``Newswatchers'' and ``Momentum traders''. The model consider three assumptions: $1)$ ``Newswatchers'' realize part of the public information and privately adjust their models, which are only based on macroeconomic and company specific forecasts. $2)$ ``Momentum traders'' only trade on past price performance. $3)$ Private, rather than public, information diffuses gradually, since each agent has a different time frame to adjust their models. These assumptions about market agents are enough to model the relationship between news and long-term trends or short-term reversals.} \cite{Hong1999} on stock's momentum/reversal proposes that investors underreact to news, causing slow price drifts, and overreact to price shocks not accompanied by news, hence inducing reversals. This theoretical predicated behaviour between price and news was systematically estimated and supported in \cite{Chan2003,Boudoukh2013} using financial media headlines and in \cite{Antoniou2013}  using the Consumer Confidence Index\textsuperscript{\textregistered} published by \href{https://www.conference-board.org/}{The Conference Board} \cite{cci2011}. Similarly, \cite{Tetlock2007} uses the \textit{Harvard IV-4} sentiment lexicon to count the occurrence of words with positive and negative connotation of the \textit{Wall Street Journal} showing that negative sentiment is a good predictor of price returns and trading volumes.  

%The textual features are represented as a vector of the joint probability distribution over topics and sentiments and a SVM model is trained using market data (price) and the texts from Yahoo message board. %The Message Board dataset of this works is relatively small, containing only one year of messages (from July, 2012 to July, 2013).
%A corpus of financial media news, similar to the one proposed in this work, is employed in 

%From the works mentioned above we can see that collecting textual data and automatically applying NLP techniques aiming to extract the meaning of textual content turned into a key tool for assessing sensible Financial Economics theories. Further, we note the breadth of datasets employed ranging from Message boards, Twitter, Media news and low-frequency business reports. 

Accurate models for forecasting both price returns and volatility are equally important in the financial domain. Volatility measures how wildly the asset is expected to oscillate in a given time period and is related to the second moment of the price return distribution. In general terms, forecasting price returns is relevant to take speculative positions. The volatility, on the other hand, measures the risk of these positions. On a daily basis, financial institutions need to assess the short-term risk\footnote{Usually, this risk is the conditional volatility for the next trading day} of their portfolios. Measuring the risk is essential in many aspects. It is imperative for regulatory capital disclosures required by banking supervision bodies. Moreover, it is useful to dynamically adjust position sizing accordingly to market conditions, thus, maintaining the risk within reasonable levels.
 
Although, it is crucial to predict the short-term volatility from the financial markets application perspective, much of the current NLP research on volatility forecasting focus on the volatility prediction for very long-term horizons (see \cite{Kogan2009, Wang2013, Tsai2014, Nopp2015,Rekabsaz2017}). Predominately, these works are built on extensions of the bag-of-words representation that has the main drawback of not capturing word order. Financial forecasting, however, requires the ability to capture semantics that is dependent upon word order. For example, the headline \textit{``Qualcomm sues Apple for contract breach''} and \textit{``Apple sues Qualcomm  for contract breach''} trigger different responses for each stock and for the market aggregated index, however, they share the same bag-of-words representation. Additionally, these works use features from a pretrained sentiment analyis model to train the financial forecasting model. A key limitation of this process is that it requires a labelled sentiment dataset. Additionally, the error propagation is not end-to-end. %In contrast, we describe an end-to-end system, that jointly learns word representations relevant for the market data.

In this work, we fill in the gaps of volatility prediction research in the following manner:
\begin{enumerate}[1.]
\item To move away from long-horizon volatility\footnote{The long-term forecast characteristic of the works described above is explained by the fact that the 10-K reports are only released annually.} to short-term daily volatility prediction, we introduce a corpus of Reuters financial news. We compiled this corpus at individual stock level comprising the news titles (headlines) of 50 stocks in 5 diversified sectors with a total of 146,783 samples (2007--2017). We also collected daily stock prices from Yahoo Finance website for the 50 stocks. \label{itm:coprpus}
%\item Previous works made large contributions predicting the volatility after the 10-K reports are released. Inspired by studies of the financial area on market underreaction/overreaction in the presence/absence of news discussed above (\cite{Hong1999,Chan2003,Boudoukh2013}), the present work models the volatility on a daily basis using Multimodal learning techniques \cite{Baltrusaitis2017}. To this aim, our model jointly learns market data (price mode) and company news (textual mode) end-to-end for all the trading days of the time series, including days when news are not released. The proposed architecture employs two Long Short Term Memory (LSTM) \cite{Hochreiter1997} models. One to capture the semantic composition over the headline words and another to encode the price features. Both encodings are then ``fused'' (concatenated) in an \textit{early fusion} fashion and trained end-to-end.
\item We propose an end-to-end multimodal model that jointly learns from daily stock price and company news.
%\item From a multimodal perspective, we are particularly interested in jointly learning and in comparing the results with unimodal price only models. In the context of multimodal learning theory it is equivalent to ask whether the textual mode is complementary or redundant  for the short-term volatility prediction problem, i.e. whether the textual mode is able to capture information not available in the price mode.
%\item In the context of multimodal learning theory, we investigate if the textual mode is complementary or redundant for the short-term volatility prediction problem, i.e. whether the textual mode is able to capture information not available in the market data mode.
\item We investigate if the textual mode is complementary or redundant for the short-term volatility prediction problem. Our results indicate that textual mode is complementary and improves the forecasting accuracy.
\item We contribute to the Universal Sentence Representation works in \cite{Conneau2017,Mou2016,Howard2018} by comparing how transferable are the representations learnt in two different NLP tasks to the specific problem of volatility forecasting.\label{itm:tranfer}
\item We propose a hierarchical \textit{news relevance attention} mechanism that can effectively select the most relevant headline news from the large amount of news released in a given day.
%\item Finally, we propose a hierarchical neural attention mechanism at the headline level. This attention aims to learn from the large amount of news released in a given day what are the most relevant ones. We named it the \textit{News Relevance Attention}.\label{itm:attention}
%(\cite{Raffel2015,Liu2016})
\end{enumerate}

%The remainder of the paper is organized as follows. 
%Additionally, the document representations do not take into account the order of the words, i.e. they use term-level features alone.  

%The question whether news impact stock markets and the efficiency of the market to rapidly absorb any fact or event has been studied in the context of stock price momentum and reversals for many decades. 
%\footnote{The statistics are based on counting queries related to the words "finance" and "text mining"/"sentiment analysis" from \href{https://www.elsevier.com/en-gb/solutions/scopus}{Scopus Database.}}.

\section{Related work}

Previous work in \citep{Kogan2009} incorporates sections of the ``Form 10-K''\footnote{Companies with listed stocks are enforced by the U.S. Securities and Exchange Commission (SEC) to file ``Form 10-K'' reports on an annual/quarterly basis. These forms provide an overview of the company's business and financial health. A 10-K form example can be found \href{https://www.sec.gov/Archives/edgar/data/320193/000032019318000145/a10-k20189292018.htm}{here}} to predict the volatility twelve months after the report is released. They train a  Support Vector Regression model on top of sparse representation (bag-of-words) with standard term weighting  (e.g. Term-Frequency). This work was extended in \cite{Wang2013,Tsai2014,Nopp2015,Rekabsaz2017} by employing the \textit{Loughran-McDonald Sentiment Word Lists} \cite{Loughran2011}, which contain three lists where words are grouped by their sentiments (positive, negative and neutral). In all these works, the textual representation is engineered using the following steps: 1) For each sentiment group, the list is expanded by retrieving 20 most similar words for each word using Word2Vec word embeddings \cite{MikolovNIPS2013}. 2) Finally, each 10-K document is represented using the expanded lists of words. The weight of each word in this sparse representation is defined using Information Retrieval (IR) methods such as \textit{term-frequency} (tf) and \textit{term-frequency with inverted document frequency} (tfidf). Particularly, \cite{Rekabsaz2017} shows that results can be improved using enhanced IR methods and projecting each sparse feature into a dense space using Principal Component Analysis (PCA).

The works described above (\cite{Wang2013,Tsai2014,Nopp2015,Rekabsaz2017}) target long-horizon volatility predictions (one year or quarterly \cite{Rekabsaz2017}). In particular, \cite{Rekabsaz2017} and \cite{Nopp2015} uses market data (price) features along with the textual representation of the 10-K reports. These existing works that employ multi-modal learning \cite{Baltrusaitis2017} are based on a \textit{late fusion}\footnote{In the \textit{late fusion} setup, text and price features are trained independently and a meta model is used in a later stage to discriminate how to weight the contribution of each mode.} approach. For example, {stacking} ensembles to take into account the price and text predictions \cite{Rekabsaz2017}. In contrast, our end-to-end trained model can learn the joint distribution of both price and text.

Predicting the price direction rather than the volatility was the focus in \cite{Bollen2011}. They extracted sentiment words from Twitter posts to build a time series of collective Profile of Mood States (POMS). Their results show that collective mood accurately predicts the direction of Down Jones stock index (86.7\% accuracy). In \cite{Schumaker2009} handcrafted text representations including term count, noun-phrase tags and extracted named entities are employed for predicting stock market direction using Support Vector Machine (SVM). An extension of Latent Dirichlet Allocation (LDA) is proposed in \cite{Nguyen2015} to learn a joint latent space of topics and sentiments.

Our deep learning models bear a close resemblance to works focused on directional price forecasting \cite{Ding2015,Pinheiro2017}. In \cite{Ding2015}, headline news are processed using \textit{Stanford OpenIE} to generate triples that are fed into a Neural Tensor Network to create the final headline representation. In \cite{Pinheiro2017}, a character-level embedding is pre-trained in an unsupervised manner. The character embedding is used as input to a sequence model to learn the headline representation. % This corpus is also  composed of financial media news. However, it has headlines only until 2013.
Particularly, both works average all headline representations in a given day, rather than attempting to weight the most relevant ones. In this work, we propose a neural attention mechanism to capture the \textit{News Relevance} and provide experimental evidence that it is a key component of the end-to-end learning process. Our attention extends the previous deep learning methods from \cite{Ding2015, Pinheiro2017}.
%In addition, the deep learning models are unimodal (only text mode)

%To illustrate the importance of ``filtering'' irrelevant information, we take an example from our corpus. The news \textit{``Exxon meets green groups''} and \textit{``Honduras temporarily grabs Exxon terminals''} were released on the same day. It is reasonable to assume that the second news, as a political event, is expected to shake the markets and should receive more weight during the training process

%The corpus of financial news introduced in our work is larger than the 10-K corpus proposed in previous ones\footnote{The 10-K corpus has \~58,000 samples}.
Despite the fact that end-to-end deep learning models have attained state-of-the-art performance, the large number of parameters make them prone to overfitting. Additionally, end-to-end models are trained from scratch requiring large datasets and computational resources. Transfer learning (TL) alleviates this problem by adapting representations learnt from a different and potentially weakly related source domain to the new target domain. For example, in computer vision tasks the convolutional features learnt from ImageNet \cite{Deng2009} dataset (source domain) have been successfully transferred to multiple domain target tasks with much smaller datasets such as object classification and scene recognition \cite{Razavian2014}. %The transfered features from ImageNet even surpassed the performance of engineered image features in the target datasets \cite{Razavian2014}.

In this work, we consider TL in our experiments for two main reasons. First, it address the question whether our proposed dataset is suitable for end-to-end training since the performance of the transferred representations can be compared with end-to-end learning. Second, it is still to be investigated which dataset transfers better to the forecasting problem. Recently, the NLP community has focused on universal representations of sentences \cite{Conneau2017,Howard2018}, which are dense representations that carry the meaning of a full sentence. \cite{Conneau2017} found that transferring the sentence representation trained on the Stanford Natural Language Inference (SNLI) \cite{Bowman2015} dataset achieves state-of-the-art sentence representations to multiple NLP tasks (e.g. sentiment analysis, question-type and opinion polarity). Following \cite{Conneau2017}, in this work, we investigate the suitability of SNLI and Reuters RCV1 \cite{Lewis2004} datasets to transfer learning to the volatility forecasting task. %Moreover, we compare the results with our end-to-end models trained with fixed Glove word embeddings \cite{Pennington2014}. %The Reuters RCV1 is a dataset of financial news with annotated categories such as Merge \& Acquisitions, Legal Issue and Company's Forecasting. %The initial assumption was that for stock market prediction problem a financial domain dataset would be more effective in capturing the financial content of our corpus sentences (headlines). However, our experiments show that the sentence representations from SNLI generalize better than Reuters RCV1 yielding better transfer performance, though the TL models do not beat our end-to-end architectures.

To the best of our knowledge, the hierarchical attention mechanism at headline level, proposed in our work, has not being applied to volatility prediction so far; neither has been investigated the ability to transfer sentence encoders from source datasets to the target forecasting problem (Transfer Learning). 

\section{Our dataset}

Our corpus covers a broad range of news including news around earnings dates and complements the 10-K reports content. As an illustration, the headlines \textit{``Walmart warns that strong U.S. dollar will cost \$15B in sales''} and \textit{``Procter \& Gamble Co raises FY organic sales growth forecast after sales beat''} describe the company financial conditions and performance from the management point of view -- these are also typical content present in Section 7\footnote{The section is called ``Management's Discussion and Analysis of financial conditions and results of operations'' (MD\&A), which is the  management's forward-looking section.} of the 10-K reports.

%Nevertheless, the coverage of our corpus demanded a specific mechanism to ``filter'' news that are released at the same time period. 

In this section, we describe the steps involved in compiling our dataset of financial news at stock level, which comprises a broad range of business sectors.

\subsection{Sectors and stocks}
\label{sub:corpus_sec_stock}
The first step in compiling our corpus was to choose the constituents stocks. Our goal was to consider stocks in a broad range of sectors, aiming a diversified financial domain corpus. We found that Exchange Traded Funds (ETF) provide a mechanical way to aggregate the most relevant stocks in a given industry/sector. An ETF is a fund that owns assets, e.g. stock shares or currencies, but, unlike mutual funds are traded in stock exchanges. These ETFs are extremely liquid and track different investment themes. We decided to use \href{https://us.spdrs.com/en/strategies/sector-industry-etfs}{SPDR Setcor Funds} constituents stocks in our work since the company is the largest provider of sector funds in the United States. We included in our analysis the top 5 (five) sector ETFs by financial trading volume (as in Jan/2018). Among the most traded sectors we also filtered out the sectors that were similar to each other. For example, the Consumer Staples and Consumer Discretionary sectors are both part of the parent Consumer category. For each of the top 5 sectors we selected the top 10 holdings, which are deemed the most relevant stocks. \ref{tbl:stock_universe}, details our dataset sectors and its respective stocks.

\begin{table*}[ht]
\centering
%\begin{tabularx}{0.50\textwidth}{L{0.30}L{0.70}}
\begin{tabularx}{1.0\textwidth}{lX}
\toprule
Sector ETF & Constituent Stocks \\ \midrule
Consumer Staples (XLP) & Procter \& Gamble (PG), Coca-Cola Company (KO), PepsiCo (PEP), Walmart (WMT), Costco Wholesale Corporation (COST), CVS Health Corporation (CVS), Altria Group (MO), Walgreens Boots Alliance (WBA), Mondelez International (MDLZ), Colgate-Palmolive (CL), \\ \midrule
Energy (XLE) & Exxon-Mobil (XOM), Chevron (CVX), ConocoPhillips (COP), EOG Resources (EOG), Occidental Petroleum Corporation (OXY), Valero Energy Corporation (VLO), Halliburton Company (HAL), Schlumberger Limited (SLB), Pioneer Natural Resources (PXD), Anadarko Petroleum Corporation (APC) \\ \midrule
Utilities (XLU) & NextEra Energy (NEE), Duke Energy (DUK), The Southern Company (SO), Dominion Energy (D), Exelon Corporation (EXC), American Electric Power Company (AEP), Sempra Energy (SRE), Public Service Enterprise Group (PEG), Consolidated Edison (ED), Xcel Energy (XEL) \\ \midrule
Healthcare (XLV) & Johnson \& Johnson (JNJ), UnitedHealth Group (UNH), Pfizer (PFE),  Merck \& Co. (MRK), Medtronic (MDT), Amgen (AMGN), Abbott Laboratories (ABT), Gilead Sciences (GILD), Eli Lilly (LLY), Bristol-Myers Squibb (BMY) \\ \midrule
Financials (XLF) & Berkshire Hathaway (BRK-A), JPMorgan Chase (JPM), Bank of America Corporation (BAC), Wells Fargo (WFC), CitiBank (C), Goldman Sachs Group (GS), U.S. Bancorp (USB), Morgan Stanley (MS), American Express (AXP), PNC Financial Services Group (PNC) \\
\bottomrule
\end{tabularx}
\captionsetup{width=1.0\textwidth}
\caption{\textbf{Corpus sectors and respective constituent stocks}. For each sector we selected the top 10 stock holdings (as in January 2018). Stock codes in parentheses.}
\label{tbl:stock_universe}
\end{table*}

\subsection{Stock specific data}
We assume that an individual stock news as the one that explicitly mention the stock name or any of its \textit{surface forms} in the headline. As an illustration, in order to collect all news for the stock code PG, \textit{Procter \& Gamble} company name, we search all the headlines with any of these words: \texttt{Procter\&Gamble} \textit{OR} \texttt{Procter and Gamble} \textit{OR} \texttt{P\&G}. In this example, the first word is just the company name and the remaining words are the company surface forms.

We automatically derived the surface forms for each stock by starting with a seed of surface forms extracted from the DBpedia Knowledge Base (KB). We then applied the following procedure:
\begin{itemize}
\item Relate each company name with the KB entity unique identifier.
\item Retrieve all values of the \href{http://dbpedia.org/ontology/wikiPageRedirects}{\texttt{wikiPageRedirects}} property. The property holds the names of different pages that points to the same entity/company name. This step sets the initial seed of surface forms.
\item Manually, filter out some noisy property values. For instance, from the \href{http://dbpedia.org/page/Procter_\%26_Gamble}{Procter \& Glamble} entity page we were able to automatically extract \linebreak
\texttt{dbr:Procter\_and\_gamble} and \texttt{dbr:P\_\&\_G}, but had to manually exclude the noisy associations \linebreak \texttt{dbr:Female\_pads} and \texttt{dbr:California\_Natural}.
\end{itemize}

The result of the steps above is a dictionary of surface forms $wd_{sc}$. %This dictionary can be found in the file \texttt{surface\_forms.pkl}.

\subsection{Stock headlines}
\label{subsec:stock_headlines}
%\href{https://uk.reuters.com/resources/archive/us/}{Reuters Archive}
Our corpus is built at stock code level by collecting headlines from the \href{https://agency.reuters.com/en/products-services/products/news-archival-footage.html}{Reuters Archive}. This archive groups the headlines by date, starting from 1 January 2007. Each headline is a html link (\texttt{<a href>} tag) to the full body of the news, where the \textit{anchor text} is the headline content followed by the release time. For example, the page dated 16 Dec 2016 has the headline \href{https://www.reuters.com/article/procter-gamble-trian/procter-gamble-appoints-nelson-peltz-to-board-idUSL4N1OF5JC}{\textit{``Procter \& Gamble appoints Nelson Peltz to board}} \textit{5:26PM UTC''}.

%We draw attention to some inconsistencies between the archive date and the real date that the headline links to. Taking a random sample of 1,000 news in 2007 we found that 15\% of   
For each of the 50 stocks (5 sectors times 10 stocks per sector) selected using the criteria described in \ref{sub:corpus_sec_stock}, we retrieved all the headlines from the Reuters Archive raging from 01/01/2007 to 30/12/2017. This process takes the following steps:
\begin{itemize}
\item For a given stock code ($sc$) retrieve all surface forms $wd_{sc}$.
%\item For each day in our historical range, parse all html anchor texts (headlines) and store only the headlines content matching any word in $wd_{sc}$. For each stored headline we parse time and timezone.
\item For each day, store only the headlines content matching any word in $wd_{sc}$. For each stored headline we also store the time and timezone.
\item Convert the news date and time to Eastern Daylight Time (EDT)\footnote{The timezone of the New York Stock exchange}.
\item Categorize the news release time. We consider the following category set: \{\texttt{before market}, \texttt{during market} , \texttt{after market}, \texttt{holidays}, \texttt{weekends}\}. \texttt{during market} contains news
between 9:30AM and 4:00PM. \texttt{before market} before 9:30AM and
\texttt{after market} after 4:00PM.  
\end{itemize}

The time categories prevents any misalignment between text and stock price data\footnote{Note that changing the timezone can change the original news date.}. Moreover, it prevents data leakage and, consequently, unrealistic predictive model performance. In general, news released after 4:00PM EDT can drastically change market expectations and the returns calculated using close to close prices as in the GARCH(1,1) model (see 
\ref{eq:closing_return}). Following \cite{Antweiler2004}, to deal with news misalignment, news issued after 4:00PM (\texttt{after market}) are grouped with the pre-market (\texttt{before market}) on the following trading day.

\ref{tbl:stock_time_cat} shows the distribution of news per sector for each time category. We can see a high concentration of news released before the market opens (55\% on average). In contrast, using a corpus compiled from message boards, a large occurrence of news during market hours was found \cite{Antweiler2004}. This behaviour indicating day traders' activity. %We argue that the distribution of news is intimately related with its source.
Our corpus comprise financial news agency headlines, a content more focused on corporate events (e.g. lawsuits, merges \& acquisitions, research \& development) and on economic news (see \ref{tbl:stock_headlines_exmaples} for a sample of our dataset). These headlines are mostly factual. On the other hand, user-generated content such as Twitter and message boards (as in \cite{Antweiler2004,Sprenger2014}) tends to be more subjective.

\begin{table*}[ht]
\centering
\begin{tabularx}{0.90\textwidth}{L{0.40}C{0.20}C{0.20}C{0.20}}
\toprule
\multirow{2}{*}{Sector ETF} & before & during & after \\
& market & market & market \\ \midrule
Consumer Staples & 54\% & 31\% & 15\% \\ 
Energy & 44\% & 36\% & 20\% \\ 
Utilities & 58\% & 31\% & 11\% \\ 
Healthcare & 55\% & 28\% & 17\% \\ 
Financials & 63\% & 24\% & 13\% \\ \midrule
\textit{total} & 84,556 & 40,996 & 21231 \\
\bottomrule
\end{tabularx}
\captionsetup{width=0.90\textwidth}
\caption{\textbf{Distribution of headlines per sector according to market hours}. The majority of the 146,783 headlines are released before 9:30AM (\texttt{before market}). The category \texttt{after market} includes news released after 4:00PM EDT. We count the categories \texttt{holiday} and  \texttt{weekend} as \texttt{before market} since they impact the following working day.}
\label{tbl:stock_time_cat}
\end{table*}

U.S. macroeconomic indicators such as \textit{Retail Sales}, \textit{Jobless Claims} and GDP are mostly released around 8:30AM (one hour before the market opens). These numbers are key drivers of market activity and, as such, have a high media coverage. Specific sections of these economic reports impact several stocks and sectors. Another factor that contribute to the high activity of news outside regular trading hours are company earnings reports. These are rarely released during trading hours. Finally, before the market opens news agencies provide a summary of the international markets developments, e.g. the key facts during the Asian and Australian trading hours. All these factors contribute to the high concentration of pre-market news. 

\begin{table*}[ht]
\centering
\begin{tabularx}{0.90\textwidth}{cX}
\toprule
Date and time & Headline \\ \midrule
2011-12-13 \linebreak 00:18:39 EDT & \textit{Valero reports power outage at Port Arthur refinery} \\ \midrule
2007-04-17 \linebreak 08:54:27 EDT & \textit{Wells Fargo profit rises 11 pct on commercial loans} \\ \midrule
2017-12-14 \linebreak 14:40:31 EDT & \textit{Perrigo lines up bid for Merck's consumer health unit} \\ \midrule
2007-01-03 \linebreak 10:27:42 EDT & \textit{UPDATE 1-Bear Stearns ups Merck to outperform} \\ \midrule
2010-02-23 \linebreak 13:35:11 EDT & \textit{Exxon Mobil says remains bullish on Nigeria} \\ \midrule
2016-09-22 \linebreak 15:32:13 EDT & \textit{Texas regulators express ``deep concern'' over NextEra deal} \\ \midrule
2008-10-14 \linebreak 08:30:00 EDT & \textit{Smart For Life\(TM\) Now Available on Costco.com} \\
\bottomrule
\end{tabularx}
\captionsetup{width=0.90\textwidth}
\caption{\textbf{Random samples from our dataset.} Note the factual/objective characteristic of our corpus, where typical news do not carry any sentiment connotation.}
\label{tbl:stock_headlines_exmaples}
\end{table*}

\section{Background}
We start this section by reviewing the GARCH(1,1) model, which is a strong benchmark used to evaluate our neural model. We then review the source datasets proposed in the literature that were trained independently and transfered to our volatility prediction model. Finally, we review the general architectures of sequence modelling and attention mechanisms.
     
\subsection{GARCH model}
\label{sec:GARCH}
Financial institutions use the concept of ``Value at risk'' to measure the expected volatility of their portfolios. The widespread econometric model for volatility forecasting is the Generalized Autoregressive Conditional Heteroskedasticity (GARCH) \cite{Engle1982,Bollerslev1986}. Previous research shows that the GARCH(1,1)\footnote{The GARCH($p$,$q$) model is specified in terms of the number of lagged terms $p$ and $q$. The GARCH(1,1) specification considers only one lagged volatility ($p=1$) and shock ($q=1$) terms.} model is hard to beat. For example, \cite{Hansen2005} compared GARCH(1,1) with 330 different  econometric volatility models showing that they are not significantly better than GARCH(1,1).
Let $p_t$ be the price of an stock at the end of a trading period with closing returns $r_t$ given by
\begin{equation}
r_t = \frac{p_t}{p_{t-1}} - 1 \label{eq:closing_return}
\end{equation}
The GARCH process explicitly models the time-varying volatility of asset returns. In the GARCH(1,1) specification the returns series $r_t$ follow the process:
\begin{align}
r_t 		&= \mu + \epsilon_t \label{eq:garch_cond_mean} \\
\epsilon_t 	&= \sigma_t z_t \label{eq:garch_white_noise} \\
\sigma^2_t 	&= a_0 + a_1 \epsilon_{t-1}^2 + b_1 \sigma_{t-1}^2
\label{eq:garch_cond_variance}
\end{align}

where $\mu$ is a constant (return drift) and $z_t$ is a sequence of i.i.d. random variables with mean zero and unit variance.
It is worth noting that although the conditional mean return described in \ref{eq:garch_cond_mean} has a constant value, the conditional volatility $\sigma_t$ is time-dependent and modeled by \ref{eq:att}.

%\begin{equation}
%\sigma_u = \sqrt{a_0 / (1 - a_1 - b_1)}. \label{eq:un_var}
%\end{equation}

%As a parametric model, GARCH(1,1) has many desired properties. One of them is the mean-reverting characteristic of the volatility. Financial markets can experience periods of high and persistent volatility, e.g. during the \textit{Credit Crunch} (2008) or the \textit{Internet Bubble Burst} (2000). However, evidence shows  that the volatility tends to return to its pre-crisis levels. This volatility behavior is captured in the GARCH(1,1) model. In order to demonstrate the mean-reverting behavior of the volatility prediction using GARCH(1,1) we analyze its forecasts.%
\subsubsection{Forecasting}
The one-step ahead expected volatility forecast can be computed directly from \ref{eq:garch_cond_variance} and is given by

%We start by forecasting the conditional volatility one time period ahead. For a given time $T$ the expected volatility given all the information available at time $T$, defined as $E_T[\sigma_{T+1}^2]$,% 
\begin{equation}
E_T[\sigma_{T+1}^2] = a_0 + a_1 E_T[\epsilon^2] + b_1 E_T[\sigma_{T}^2] \label{eq:forecast_one_period}
\end{equation}
In general, the $t^{\prime}$-steps ahead expected volatility $E_T[\sigma_{T+t^{\prime}}^2]$ can be easily expressed in terms of the previous step expected volatility.
%For example, the forecast for $t^{\prime}=2$ can be written in terms of $E_T[\sigma_{T+1}^2]$ as
%\begin{equation*}
%E_T[\sigma_{T+2}^2] = a_0 + (a_1 + b_1) E_T[\sigma_{T+1}^2],\label{eq:forecast_1}
%\end{equation*}
%where we use \ref{eq:forecast_one_period} and the fact that $E_T[\epsilon_{T+1}^2] = E_T[\sigma_{T+1}^2 z_{T+1}^2] = E_T[\sigma_{T+1}^2]$ since $z_t$ has unit variance.
%A simple way to derive the forecasting asymptotic for long horizons is to rewrite the equation above in terms of the unconditional variance \ref{eq:un_var}
%\begin{equation}
%E_T[\sigma_{T+2}^2] - \sigma_u^2 = (a_1 + b_1) \left(E_T[\sigma_{T+1}^2] - \sigma_u^2\right), \label{eq:forecast_two_period}
%\end{equation}
%where $\sigma_u$ is the unconditional volatility.
It is easy to prove by induction that the forecast for any horizon can be represented in terms of the one-step ahead forecast and is given by
\begin{equation}
E_T[\sigma_{T+t^{\prime}}^2] - \sigma_u^2 = (a_1 + b_1)^{(t^{\prime} -1)} \left(E_T[\sigma_{T+1}^2] - \sigma_u^2\right)
\label{eq:forecast_recursive}
\end{equation}
where $\sigma_u$ is the \textit{unconditional volatility}:
\begin{equation}
\sigma_u = \sqrt{a_0 / (1 - a_1 - b_1)} \label{eq:un_var}
\end{equation} 
From the equation above we can see that for long horizons, i.e. $t^\prime \to \infty$, the volatility forecast in \ref{eq:forecast_recursive} converges to the unconditional volatility in \ref{eq:un_var}.
%The average time that the volatility takes to revert to its long run level can be measured in terms of the \textit{half-life of a volatility shock} and is given by $\ln(1/2)/\ln(a_1 + b_1)$. The half-life measures how long it takes to $\epsilon_{T+t^{\prime}}^2 - \sigma_u^2$ to decrease by half. %

All the works reviewed in \ref{sec:introduction} (\cite{Kogan2009,Wang2013,Tsai2014,Nopp2015,Rekabsaz2017}) consider GARCH(1,1) benchmark. However, given the long horizon of their predictions (e.g. quarterly or annual), the models are evaluated using the unconditional volatility $\sigma_u$ in \ref{eq:un_var}. In this work, we focus on the short-term volatility prediction and use the GARCH(1,1) one-day ahead conditional volatility prediction in \ref{eq:forecast_one_period} to evaluate our models.  

\subsubsection{Evaluation}
\label{sub:evalution}
%The one-day ahead volatility forecast $E_t[\sigma_{t+1}]$ is used to evaluate the daily performance of GARCH models. %
Let $\sigma_{t+1}$ denote the ex-post ``true'' daily volatility at a given time $t$. The performance on a set with $N$ daily samples can be evaluated using the standard Mean Squared Error ($MSE$) and Mean Absolute Error ($MAE$)
\begin{align}
MSE &= \frac{1}{N} \sum_{t=1}^{N} \left ( E_t[\sigma_{t+1}] - \sigma_{t+1}\right)^2 \label{eq:rmse_loss} \\
MAE &= \frac{1}{N} \sum_{t=1}^{N}\left \lvert E_t[\sigma_{t+1}] - \sigma_{t+1} \right \rvert \label{eq:mae_loss}
\end{align}

Additionally, following \cite{Andersen1998}, the models are also evaluated using the coefficient of determination $R^2$ of the regression

\begin{equation}
\sigma_{t+1} = a + b E_t[\sigma_{t+1}] + e_t
\label{eq:regression_loss}
\end{equation}
where
\begin{equation}
R^2 = 1 - \frac{\sum_{t=1}^{N}e^{2}_{t}}{\sum_{t=1}^{N}\left(E_t[\sigma_{t+1}] -  \frac{1}{N} \sum_{t=1}^{N}E_t[\sigma_{t+1}]\right)^{2}}
\end{equation}

One of the challenges in evaluating GARCH models is the fact that the ex-post volatility $\sigma_{t+1}$ is not directly observed. Apparently, the squared daily returns $r_{t+1}^{2}$ in \ref{eq:closing_return} could stand as a good proxy for the ex-post volatility. However, the squared returns yield very noisy measurements. This is a direct consequence of the term $z^t$ that connects the squared return to the latent volatility factor in \ref{eq:garch_white_noise}. %In a positional paper, \cite{Andersen1998} evaluated the GARCH(1,1) model using ex-post volatility estimators based on intraday data, rather than using squared daily returns. The work found that estimating the ex-post volatility using intraday returns provides a dramatic improvement in the GARCH(1,1) model evaluation metrics. Moreover, it explains the apparent poor forecasting performance for models evaluated using squared returns. Even in the hypothetical case that GARCH(1,1) is the true stochastic process specification, the Coefficient of determination $R^2$ is equal to $\kappa^{-1}$ where $\kappa$ is the kurtosis of the residuals $z_t$ \cite{Andersen1998}. As a consequence, in the case that $z_t$ is a Gaussian error we have $R^2$ of GARCH(1,1) bounded from above by $1/3$. For example, considering the Deutsche Mark daily exchange rate for GARCH(1,1) evaluation, \cite{Andersen1998} found $R^2$ values of $0.047$ and $0.33$ using the squared returns and intraday returns\footnote{The ex-post volatility estimator is calculated using squared returns of price data sampled every 5 minutes for a given trading day.} to estimate the volatility, respectively.%
The use of intraday prices to estimate the ex-post daily volayility was first proposed in \cite{Andersen1998}. They argue that volatility estimators using intraday prices is the proper way to evaluate the GARCH(1,1) model, as opposed to squared daily returns. For example, considering the Deutsche Mark the GARCH(1,1) model $R^2$ improves from $0.047$ (squared returns) to $0.33$ (intraday returns)\footnote{The intraday estimator is calculated using squared returns of price data sampled every 5 minutess.} \cite{Andersen1998}.

\subsubsection{Range measures to daily volatility proxy}
\label{subsub:range_vol_estimators}
It is clear from the previous section that any volatility model evaluation using the noisy squared returns as the ex-post volatility proxy will lead to very poor performance. Therefore, high-frequency intraday data is fundamental to short-term volatility performance evaluation. However, intraday data is difficult to acquire and costly. Fortunately, there are statistically efficient daily volatility estimators that only depend on the open, high, low and close prices. These price ``ranges'' are widely available. In this section, we discuss these estimators.

Let $O_t$, $H_t$, $L_t$, $C_t$ be the open, high, low and close prices of an asset in a given day $t$. Assuming that the daily price follows a geometric Brownian motion with zero drift and constant daily volatility $\sigma$, Parkinson (1980) derived the first daily volatility estimator
\begin{equation}
\widehat{\sigma_{PK,t}^2} = \frac{\ln\left(\frac{H_t}{L_t}\right)^2}{4\ln(2)} \label{eq:vol_pk}
\end{equation}
which represents the daily volatility in terms of its price range. Hence, it contains information about the price path. Given this property, it is expected that $\sigma_{PK}$ is less noisy than the volatility calculated using squared returns. 
The Parkinson's volatility estimator was extended by Garman-Klass (1980) which incorporates additional information about the opening ($O_t$) and closing ($C_t$) prices and is defined as
\begin{equation}
\widehat{\sigma_{GK,t}^{2}} = \frac{1}{2} \ln\left(\frac{H_t}{L_t}\right)^2 - (2\ln(2) - 1) \ln\left(\frac{C_t}{O_t}\right)^2 \label{eq:vol_gk}
\end{equation} 
The relative noisy of different estimators $\hat{\sigma}$ can be measured in terms of its relative efficiency to the daily volatility $\sigma$ and is defined as
\begin{equation}
e\left(\widehat{\sigma^{2}}, \sigma^2\right) \equiv \frac{Var[\sigma^2]}{Var[\widehat{\sigma^{2}}]}
\end{equation}
where $Var[\cdot]$ is the variance operator.
It follows directly from \ref{eq:garch_white_noise} that the squared return has efficiency 1 and therefore, very noisy. \cite{Molnar2012} reports Parkinson ($\widehat{\sigma_{PK,t}^2}$)
volatility estimator has  4.9 relative efficiency and Garman-Klass ($\widehat{\sigma_{GK,t}^2}$) 7.4. Additionally, all the described estimators are unbiased.

Many alternative estimators to daily volatility have been proposed in the literature. However, experiments in \cite{Molnar2012} rate the Garman-Klass volatility estimator as the best volatility estimator based only on open, high, low and close prices. In this work, we train our models to predict the state-of-the-art Garman-Klass estimator. Moreover, we evaluate our models and GARCH(1,1) using the metrics described in \ref{sub:evalution}, but with the appropriate volatility proxies, i.e. Parkinson and Garman-Klass estimators. %The evaluations using the very noisy squared returns are also reported and should be taken with care since it is not a suitable proxy for the daily volatility. We found the same apparent poor performance\footnote{$R^2$ around 0.10 for squared returns and much lower than ~0.35 for the range estimators.} reported by other authors \cite{Molnar2012,Andersen1998} when using squared returns as a proxy to daily volatility.

\subsection{Transfer Learning from other source domains}
\label{sec:transfer_learning}
Vector representations of words, also known as Word embeddings \cite{MikolovNIPS2013,Pennington2014}, that represent a word as a dense vector has become the standard building blocks of almost all NLP tasks. These embeddings are trained on large unlabeled corpus and are able to capture context and similarity among words.

Some attempts have been made to learn vector representations of a full sentence, rather than only a single word, using unsupervised approaches similar in nature to word embeddings. Recently, \cite{Conneau2017} showed state-of-the-art performance when a sentence encoder is trained end-to-end on a supervised source task and transferred to other target tasks. Inspired by this work, we investigate the performance of sentence encoders trained on the Text categorization and Natural Language Inference (NLI) tasks and use these encoders in our main short-term volatility prediction task.

A generic sentence encoder $S_e$ receives the sentence words as input and returns a vector representing the sentence. This can be expressed as a mapping
\begin{equation}
S_e \colon \mathbb{R}^{T^{S} \times d_w} \to \mathbb{R}^{d_S}
\label{eq:set_encoder_map}
\end{equation}
from a variable size sequence of words to a sentence vector $S$ of fixed-size $d_S$, where $T^{S}$ is the sentence number of words and $d_w$ is the pre-trained word embedding dimension.

In the following sections, we describe the datasets and architectures to train the sentence encoders of the auxiliary transfer learning tasks.

\subsubsection{Reuters RCV1}
\label{subsec:reuters_rcv1}
The Reuters Corpus Volume I (RCV1) is corpus containing 806,791 news articles in the English language collected from 20/08/1996 to 19/08/1997 \cite{Lewis2004}. The topic of each news was human-annotated using a hierarchical structure. At the top of the hierarchy, lies the coarse-grained categories: CCAT (Corporate), ECAT (Economics), GCAT (Government), and MCAT (Markets). A news article can be assigned to more than one category meaning that the text categorization task is mutilabel. Each news is stored in a separate XML file. \ref{lst:rcv1_xml_example} shows the typical structure of an article.

%\noindent\begin{minipage}{.45\textwidth}
\begin{lstlisting}[basicstyle=\footnotesize, language=XML,breaklines=true,breakatwhitespace=false,frame=single,caption={\textbf{RCV1 dataset article example}. For brevity's sake, we only show the markup consumed in our models. This headline has root categories CCAT (Corporate/Industrial) and MCAT (Markets) with direct children categories C13 (REGULATION/POLICY), C31 (MARKETS/MARKETING) and M14 (COMMODITY MARKETS). The last category M141 (SOFT COMMODITIES) is a children of M14 and describes the commodity market type.},captionpos=b,label={lst:rcv1_xml_example}]
<?xml version="1.0" encoding="iso-8859-1" ?>
<newsitem itemid="6159" id="root" date="1996-08-21" xml:lang="en">
<headline>Colombia raises internal coffee price.</headline>
<dateline>BOGOTA 1996-08-21</dateline>
<copyright>(c) Reuters Limited 1996</copyright>
<metadata>
<codes class="bip:topics:1.0">
  <code code="C13">
    <editdetail attribution="Reuters BIP Coding Group" action="confirmed" date="1996-08-21"/>
  </code>
  <code code="C31">
    <editdetail attribution="Reuters BIP Coding Group" action="confirmed" date="1996-08-21"/>
  </code>
  <code code="CCAT">
    <editdetail attribution="Reuters BIP Coding Group" action="confirmed" date="1996-08-21"/>
  </code>
  <code code="M14">
    <editdetail attribution="Reuters BIP Coding Group" action="confirmed" date="1996-08-21"/>
  </code>
  <code code="M141">
    <editdetail attribution="Reuters BIP Coding Group" action="confirmed" date="1996-08-21"/>
  </code>
  <code code="MCAT">
    <editdetail attribution="Reuters BIP Coding Group" action="confirmed" date="1996-08-21"/>
  </code>
</codes>
</metadata>
</newsitem>
\end{lstlisting}
%\end{minipage}\hfill

The RCV1 dataset is not released with a standard train, validation, test split. In this work, we separated 15\% of samples as a test set for evaluation purposes. The remaining samples were further split leaving 70\% and 15\% for training and validation, respectively.

Regarding the categories distribution, we found that, from the original 126 categories, 23 categories were never assigned to any news; therefore, were disregarded. From the 103 classes left we found a high imbalance among the labels with a large number of underrepresented categories having less than 12 samples. The very low number of samples for these minority classes brings a great challenge to discriminate the very fine-grained categories. Aiming to alleviate this problem, we grouped into a same class all categories below the second hierarchical level. For example, given the root node CCAT (Corporate) we grouped C151 (ACCOUNTS/EARNINGS), C1511 (ANNUAL RESULTS) and C152 (COMMENT/FORECASTS) into the direct child node C15 (PERFORMANCE). Using this procedure the original 103 categories where reduced to 55. One of the benefits of this procedure was that the less represented classes end up having around thousand samples compared with only 12 samples in the original dataset.

\begin{figure*}[ht]
\centering
\includegraphics[width=0.85\textwidth]{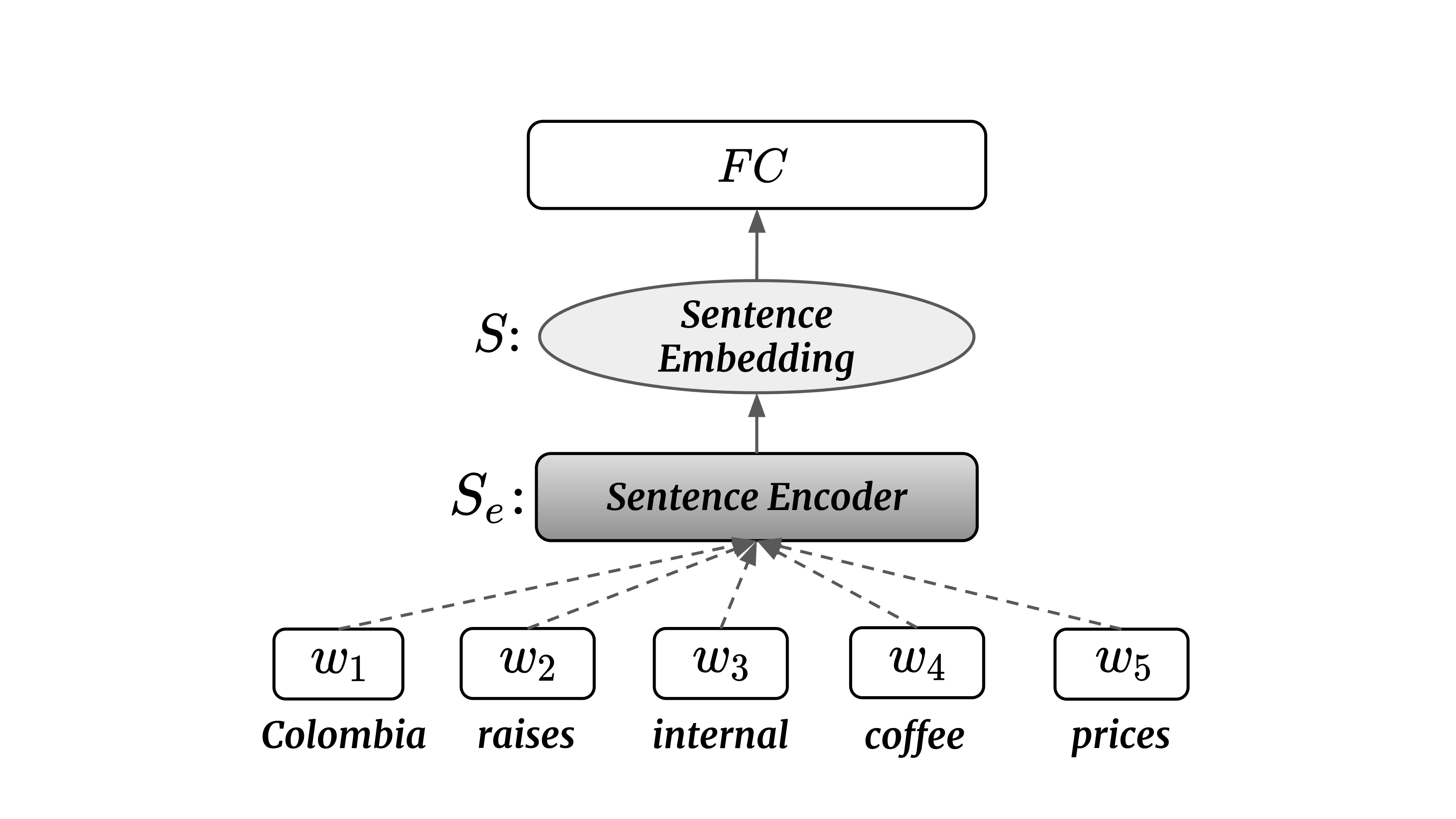}
\captionsetup{width=0.85\textwidth}
\caption{\textbf{RCV1 text categorization architecture}. The sentence encoder $S_e$ maps word emebddings $w_i$ to a sentence vector $S$ and the last FC layer has a sigmoid activation function.}
\label{fig:rcv1_arch}
\end{figure*}

\ref{fig:rcv1_arch}, shows the architecture for the end-to-end text categorization task. On the bottom of the architecture $S_e$ receives word embeddings and outputs a sentence vector $S$. The $S$ vector pass through a fully connected (FC) layer with sigmoid activation function that outputs a vector $\hat{y} \in \mathbb{R}^{55}$ with each element $\hat{y}_j \in [0,1]$.

The architecture described above is trained under the assumption that each category is independent but not mutually exclusive since a sample can have more than one category assigned (multilabel classification). The loss per sample is the average \textit{log loss} across all labels: 
\begin{equation}
\mathcal{L}(\hat{y}, y) = - \sum_{i=1}^{55}\left( y_i \log (\hat{y}_i) + (1-y_{i}) \log (1-\hat{y}_{i}) \right)
\end{equation}
where the index $i$ runs over the elements of the predicted and true vectors. 

Given the high categories imbalance, during the training we monitor the $F_1$ metric of the validation set and choose the model with the highest value.

\subsubsection{SNLI dataset}
\label{subsec:stanford_nli}
Stanford Natural Language Inference (SNLI) dataset \cite{Bowman2015} consist of 570,000 pairs of sentences. Each pair has a premise and a hypothesis, manually labeled with one of the three labels: \textit{entailment}, \textit{contradiction}, or \textit{neutral}. The SNLI has many desired properties. The labels are equally balanced, as opposed to the RCV1 dataset. Additionally, language inference is a complex task that requires a deeper understanding of the sentence meaning making this dataset suitable for learning supervised sentence encoders that generalize well to other tasks \cite{Conneau2017}. \ref{tbl:snli_exmaples}, shows examples of SNLI dataset sentence pairs and its respective labels.

\begin{table*}[ht]
\centering
\begin{tabularx}{0.90\textwidth}{XXc}
\toprule
Premise & Hypothesis & Label \\ \midrule
Children smiling and waving at camera. & There are children present. & e \\ \midrule
Two blond women are hugging one another. & Some women are hugging on vacation. & n \\ \midrule
A farmer fertilizing his garden with manure with a horse and wagon. & The man is fertilizering his garden. & e \\ \midrule 
The furry brown dog is swimming in the ocean. & A dog is running around the yard. & c \\ \midrule
A dog drops a red disc on a beach. & a dog catch the ball on a beach. & c \\ \midrule
Several armed forces officers and civilians are standing around a children's playground. & Civilians and armed forces officers trade insults at a playground. & n \\
\bottomrule
\end{tabularx}
\captionsetup{width=0.90\textwidth}
\caption{\textbf{Stanford NLI (SNLI) dataset examples.} Natural language sentence pairs are labelled with entailment (e), contradiction (c), or neutral (n).}
\label{tbl:snli_exmaples}
\end{table*}

In order to learn sentence encoders that can be transfered to other tasks unambiguously, we consider a neural network architecture for the sentence encoder with shared parameters between the premise and hypothesis pairs as in \cite{Conneau2017}.

\begin{figure*}[ht]
\centering
\includegraphics[width=0.85\textwidth]{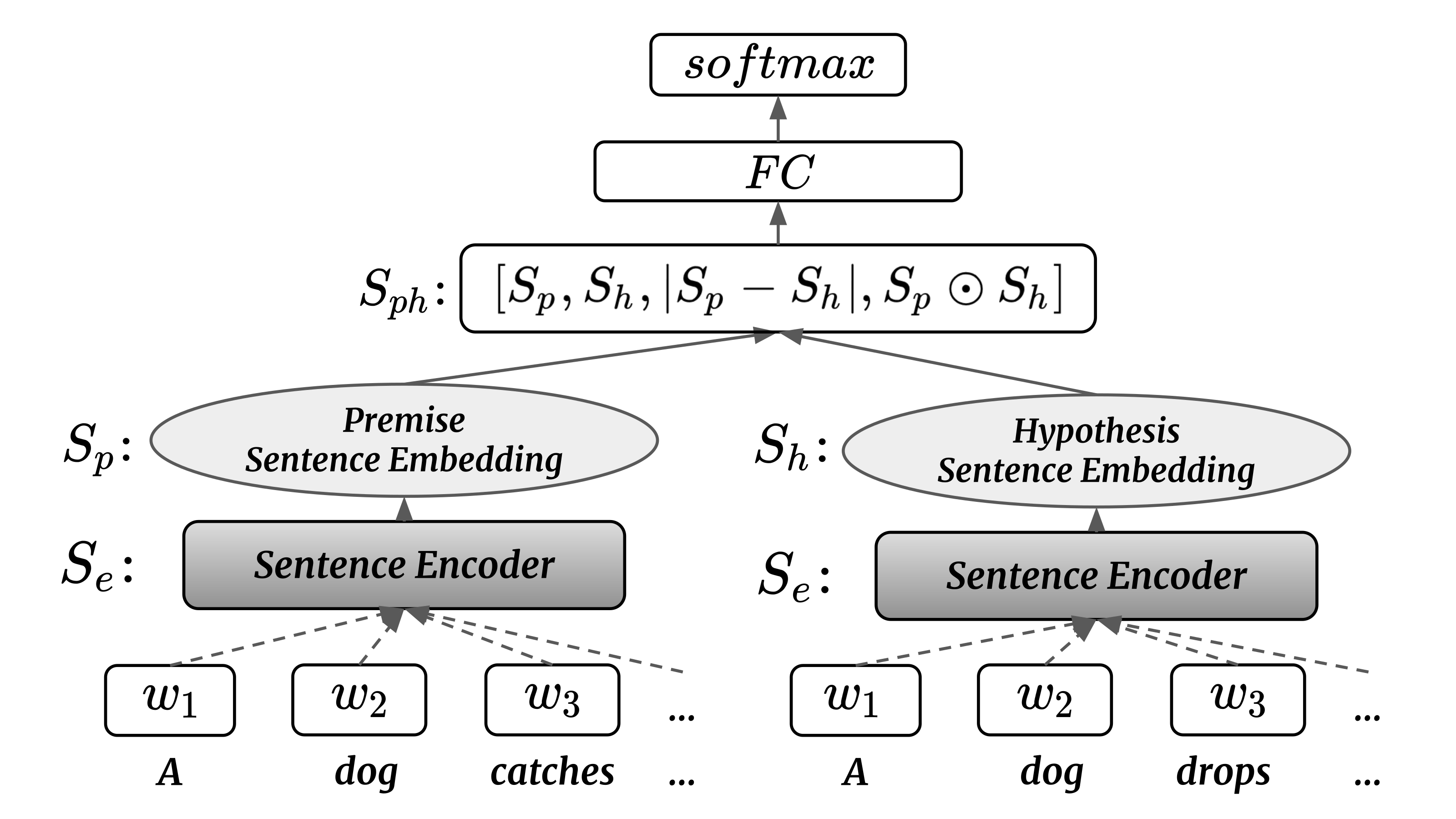}
\captionsetup{width=0.85\textwidth}
\caption{\textbf{Natural Language Inference task architecture}. Note that the sentence encoder $S_e$ is shared between the premise and hypothesis pair. The FC layer learns the representation of the sentence pair and the final Softmax layer asserts the output of the 3 possible labels, i.e.  [\textit{entailment}, \textit{contradiction}, \textit{neutral}], sums to one.}
\label{fig:snli_arch}
\end{figure*}

\ref{fig:snli_arch}, describes the neural network architecture. After each  premise and hypothesis is encoded into $S_p$ and $S_h$, respectively, we have a fusion layer. This layer has no trainable weights and just concatenate each sentence embedding. Following \cite{Conneau2017}, we add two more matching methods: the absolute difference $\vert S_p - S_h \vert $ and the element-wise $S_p \odot S_h$. Finally, in order to learn the pair representation, $S_ph$ is feed into and FC layer with rectified linear unit (ReLU) activation function, which is expressed as $f(x) = \log(1 + e^x)$. The last softmax layer outputs the probability of each class.

Finally, the NLI classifier weights are optimized in order to minimize the categorical log loss per sample
\begin{equation}
\mathcal{L}(\hat{y}, y) = - \sum_{j=1}^{3}y_i \log (\hat{y}_i)
\end{equation}
During the training, we monitor the validation set accuracy and choose the model with the highest metric value.

\subsection{Sequence Models}
\label{sec:sequence_learning}
We start this section by reviewing the Recurrent Neural Network (RNN) architecture and its application to encode a sequence of words.

RNN's are capable of handling variable-length sequences, this being a direct consequence of its recurrent cell, which shares the same parameters across all sequence elements. In this work, we adopt the Long Short-Term Memory (LSTM) cell \cite{Hochreiter1997} with forget gates $f_t$ \cite{Gers2000}. The LSTM cell is endowed with a memory state that can learn representations that depend on the order of the words in a sentence. This makes LSTM more fit to find relations that could not be captured using standard bag-of-words representations.

\begin{figure}[!ht]
\centering
\includegraphics[width=0.50\textwidth]{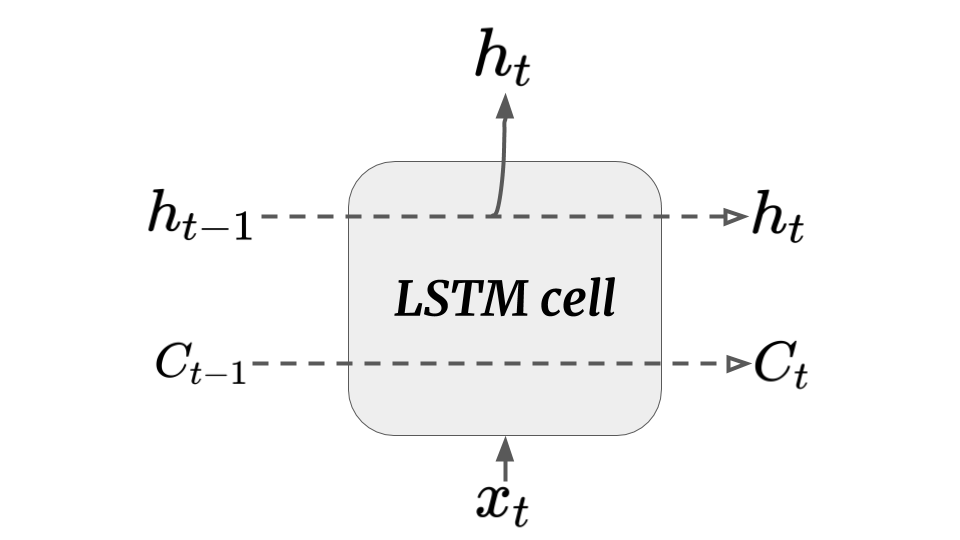}
\caption{\textbf{Schematic view of a LSTM cell}. The observed state $x_t$ is combined with previous memory and hidden states to output a hidden state $h_t$. The memory state $C_t$ is an internal state; therefore, not part of the output representation. An LSTM network is trained by looping its shared cell across all sequence length.}
\label{fig:lstm_cell}
\end{figure}

Let $x_1, x_2, \cdots ,  x_T$ be a series of observations of length $T$, where $x_t \in \mathbb{R}^{d_w}$. In general terms, the LSTM cell receives a previous hidden state $h_{t-1}$ that is combined with the current observation $x_t$ and a memory state $C_t$ to output a new hidden state $h_t$. This internal memory state $C_{t}$ is updated depending on its previous state and three modulating gates: input, forget, and output.
Formally, for each step $t$ the updating process goes as follows (see \ref{fig:lstm_cell} for a high level schematic view):
First, we calculate the input $i_t$, forget $f_t$, and output $o_t$ gates:
\begin{align}
i_t &= \sigma_s\left(W_i x_t + U_i h_{t-1} + b_i\right) \\
f_t &= \sigma_s\left(W_f x_t + U_f h_{t-1} + b_f\right) \\
o_t &= \sigma_s\left(W_o x_t + U_o h_{t-1} + b_o\right)
\label{eq:input_forget_gates}
\end{align}
where $\sigma_s$ is the sigmoid activation.
Second, a candidate memory state $\widetilde{C}_t$ is generated:
\begin{equation}
\widetilde{C}_t = \tanh\left(W_c x_t + U_c h_{t-1} + b_c\right)
\end{equation}
Now we are in a position to set the final memory state $C_t$. Its value is modulated based on the input and forget gates of \ref{eq:input_forget_gates} and is given by:
\begin{equation}
C_t = i_t \odot \widetilde{C}_t + f_t \odot C_{t-1} 
\end{equation}
Finally, based on the memory state and output gate of \ref{eq:input_forget_gates}, we have the output hidden state
\begin{equation}
h_t = o_t \odot \tanh\left(C_t\right)
\end{equation}

Regarding the trainable weights, let $n$ be the LSTM cell number of units. It follows that $W$'s and $U$'s matrices of the affine transformations have ${n \times d_w}$ and ${n \times n}$ dimensions, respectively. Its bias terms $b$'s are vectors of size $n$. Consequently, the total number of parameters is $4 (n d_w + n^2 + n)$ and does not depend on the sequence number of time steps $T$.

We see that the LSTM networks are able to capture temporal dependencies in sequences of arbitrary length. One straightforward application is to model the Sentence encoder discussed in \ref{sec:transfer_learning}, which outputs a sentence vector representation using its words as input. 

Given a sequence of words $\left\{w_t\right\}_{t=1}^{T}$ we aim to learn the words hidden state $\left\{h_t\right\}_{t=1}^{T}$ in a way that each word captures the influence of its past and future words. The Bidirectional LSTM (BiLSTM) proposed in \cite{Schuster1997} is an LSTM that ``reads'' a sentence, or any sequence in general, from the beginning to the end (forward) and the other way around (backward). The new state $h_t$ is the concatenation
\begin{equation}
h_t = [\vec{h_t}, \cev{h_t}]
\label{eq:h_t_concat}
\end{equation}
where
\begin{align}
\vec{h_t} &= \text{LSTM}\left(w_1, \cdots, w_T\right) \\
\cev{h_t} &= \text{LSTM}\left(w_T, \cdots, w_1\right) \\
\end{align}     

Because sentences have different lengths, we need to convert the $T$ concatenated hidden states of the BiLSTM into a fixed-length sentence representation. One straightforward operation is to apply any form of pooling. Attention mechanism is an alternative approach where the sentence is represented as an weighted average of hidden states where the weights are learnt end-to-end.

In the next sections we describe the sentence encoders using pooling and attention layers.

\subsubsection{BiLSTM max-pooling}
\label{subsec:bilstm_max_pool}
\begin{figure*}[ht]
\centering
\includegraphics[width=0.85\textwidth]{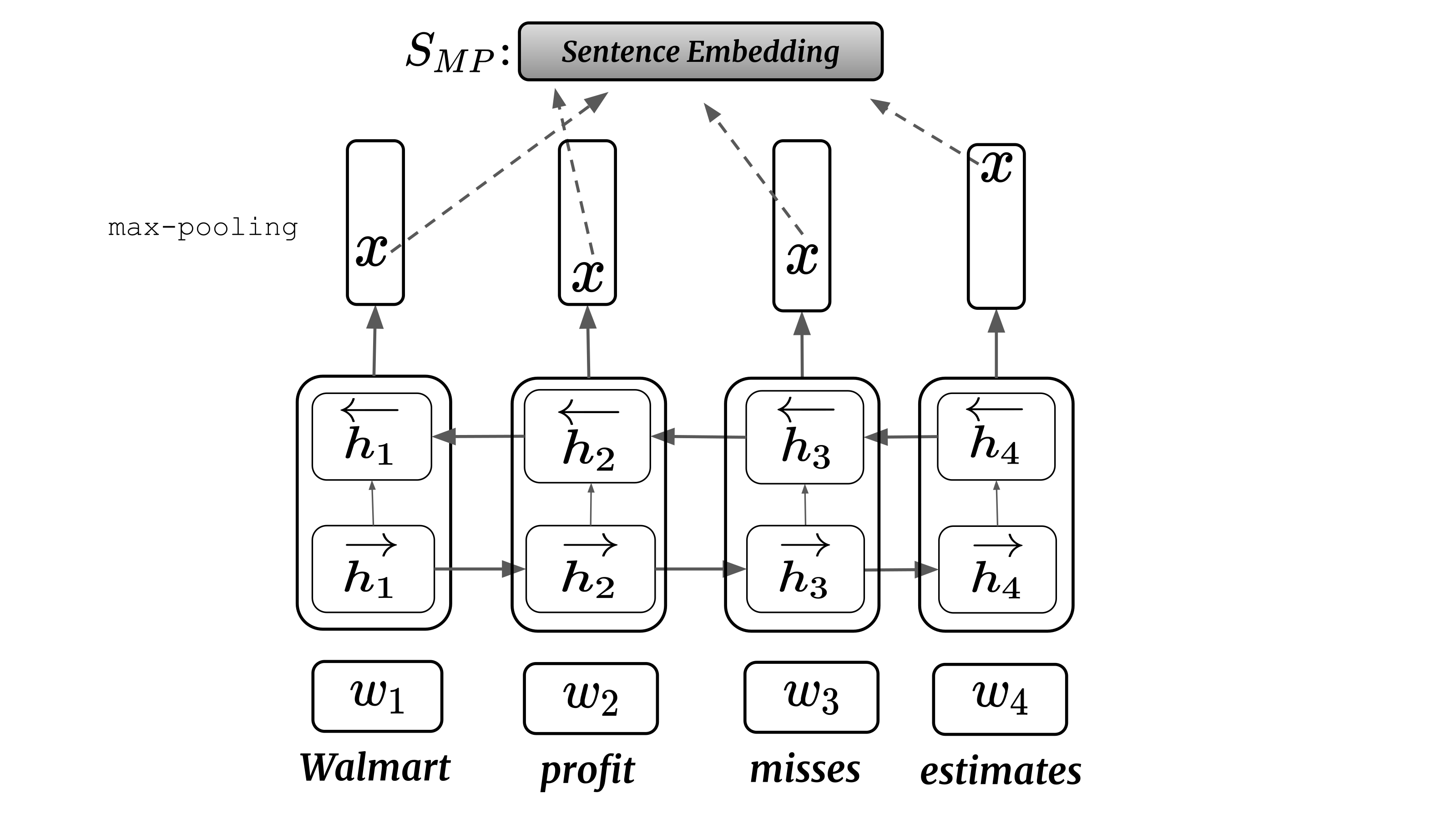}
\captionsetup{width=0.85\textwidth}
\caption{\textbf{BiLSTM max-pooling}. The network performs a polling operation on top of each word hidden state.}
\label{fig:bilstm_max_pool}
\end{figure*}

The max-pooling layer aims to extract the most salient word features all over the sentence. Formally, it outputs a sentence vector representation $S_{MP} \in \mathbb{R}^{2n}$ such that
\begin{equation}
S_{MP} = \max_{t=1}^{T} h_t
\end{equation} 
where $h_t$ is defined in \ref{eq:h_t_concat} and the $\max$ operator is applied over the time steps dimension. \ref{fig:bilstm_max_pool} illustrates the BiLSTM max-pooling (MP) sentence encoder.

The efficacy of the max-pooling layer was assessed in many NLP studies. \cite{Lai2015} employed a max-pooling layer on top of word representations and argues that it performs better than mean pooling. Experimental results in \cite{Conneau2017} show that among three types of pooling (max, mean and last\footnote{The ``last'' polling is a simple operator that takes only the last element of the $T$ hidden states to represent a sentence.}) the max-pooling provides the most universal sentence representations in terms of transferring performance to other tasks. Grounded on these studies, in this work, we choose the BiLSTM max-pooling as our pooling layer of choice.  

\subsubsection{BiLSTM attention}
\label{subsec:bilstm_inner_att}
Attention mechanisms were introduced in the deep learning literature to overcome some simplifications imposed by pooling operators. When we humans read a sentence, we are able to spot its most relevant parts in a given context and disregard information that is redundant or misleading. The attention model aims to mimic this behaviour.

Attention layers were proposed for different NLP tasks. For example, NLI, with cross-attention between premise and hypothesis, Question \& Answering and Machine Translation (MT). Specifically in the Machine Translation task, each word in the target sentence learns to attend the relevant words of the source sentence in order to generate the sentence translation.

A sentence encoder with attention (or self-attentive) \cite{Li2016,Liu2016, Lin2017} assigns different weights to the own words of the sentence; therefore, converting the hidden states into a single sentence vector representation. 

Considering the word hidden vectors set $\{h_1, \cdots, h_T\}$ where $h_t \in \mathbb{R}^n$, the attention mechanism is defined by the equations:
\begin{align}
\tilde{h}_t &= \sigma \left(W h_t + b \right) \\
\alpha_{t} &= \frac{\exp ({v^{\intercal} \cdot \tilde{h}_t} )}{\sum_{t} \exp ({v \cdot \tilde{h}_t})}  \\
S_{A_w} &= \sum_{t} \alpha_{t} h_t
\label{eq:att}
\end{align}
where $W \in \mathbb{R}^{d_a \times n}$, $b \in \mathbb{R}^{d_a \times 1}$, and $v \in \mathbb{R}^{d_a \times 1}$ are trainable parameters.

\begin{figure*}[ht]
\centering
\includegraphics[width=0.85\textwidth]{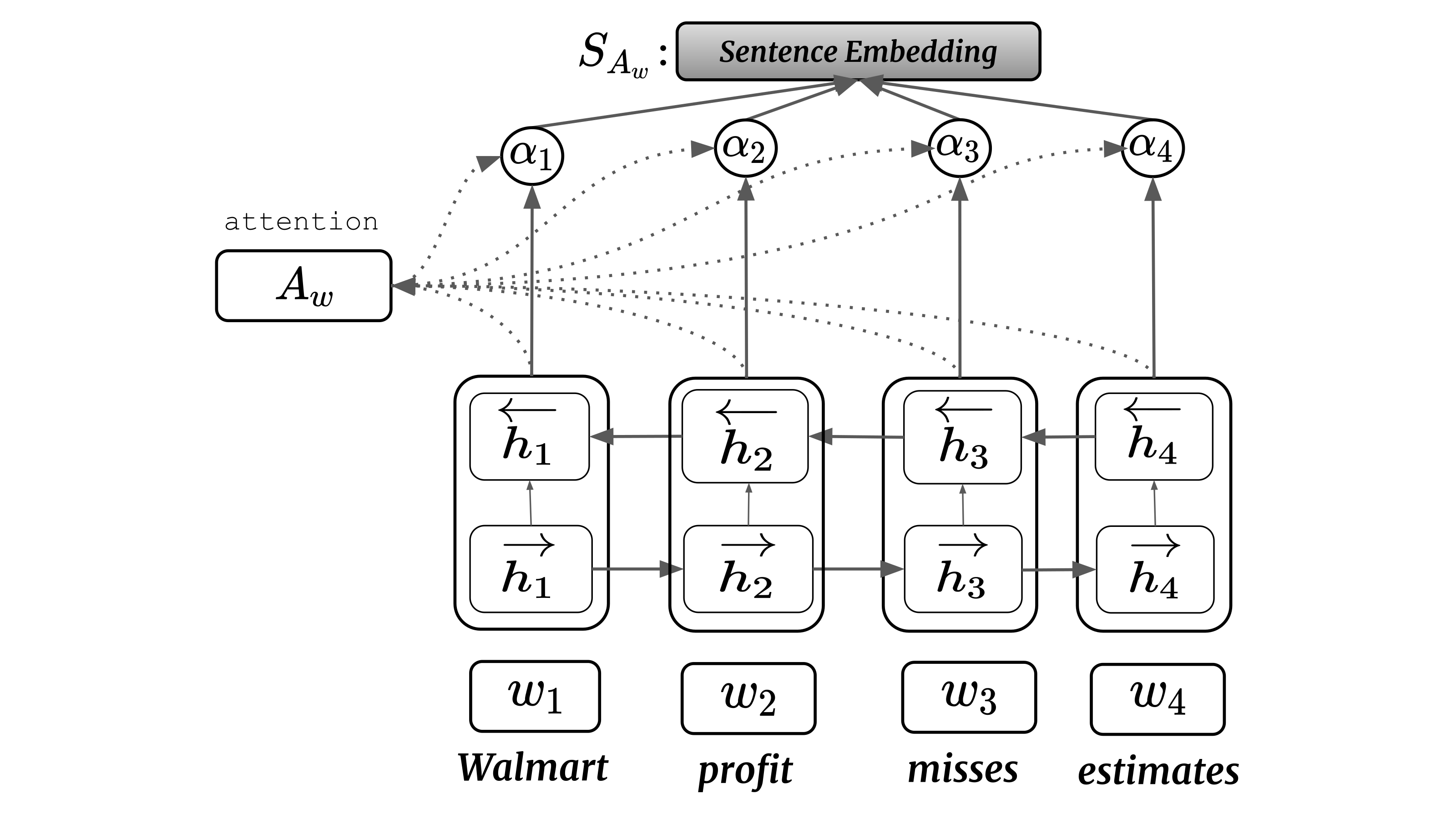}
\captionsetup{width=0.85\textwidth}
\caption{\textbf{BiLSTM attention}. The specific example encodes a headline from our corpus.}
\label{fig:bilstm_inner_att}
\end{figure*}

We can see that the sentence representation $S_{A_w}$ is a weighted average of the hidden states. \ref{fig:bilstm_inner_att} provides a schematic view of the BiLSTM attention, where we can account the attention described in \ref{eq:att} as a two layer model with a dense layer ($d_a$ units) followed by another dense that predicts $\alpha_t$ (single unit).  

\section{Methodology}
In this section, we first introduce our problem in a deep multimodal learning framework. We then present our neural architecture, which is able to address the problems of news relevance and novelty. Finally, we review the methods applied to learn commonalities between stocks (global features).
    
\subsection{Problem statement}
Our problem is to predict the daily stock volatility. As discussed in \ref{subsub:range_vol_estimators}, the Gaman-Klass estimator $\widehat{\sigma_{GK,t}}$ in \ref{eq:vol_gk} is a very efficient short-term volatility proxy, thus, it is adopted as our target variable.

%Our model is multimodal with the price features on a given day $t$ are defined without rely on any feature engineering. It is expressed as:
%\begin{equation}
%\vec{r_{t^\prime}} = \left[\frac{O_{t^\prime}}{C_{t^\prime-1} } - 1, \frac{H_{t^\prime}}{C_{t^\prime-1} } - 1, \frac{L_{t^\prime}}{C_{t^\prime-1} } - 1, \frac{C_{t^\prime}}{C_{t^\prime-1} } - 1 \right]
%\end{equation}

%Our goal is to learn a mapping between the next day volatility $\sigma_{t+1}$ and historical multimodal data available up to day $t$. To this aim, we use a \textit{sliding window} approach with window size $T$. That is, on a given day $t$ and stock $sc$ we retrieve prices and corpus data from $t-T$ to $t$. Next, we append to a vector the sample of historical prices $P_t$ and the sample corpus headlines $N_t$. Then the window is moved ahead one day $t+1$ and the same process is repeated until the last date of our series.
Our goal is to learn a mapping between the next day volatility $\sigma_{t+1}$ and historical multimodal data available up to day $t$. To this aim, we use a \textit{sliding window} approach with window size $T$.
That is, for each stock $sc$ a sample on day $t$ is expressed as a sequence of historical prices $P^{sc}_t$ and corpus headlines $N^{sc}_t$. The price sequence is a vector of Daily Prices (DP) and expressed as   
\begin{equation}
P^{sc}_t = \left[DP^{sc}_{t-T}, DP^{sc}_{t-T+1}, \cdots, DP^{sc}_t \right]
\end{equation}
where $DP^{sc}_{t^{\prime}}$ is a vector of price features. In order to avoid task-specific feature engineering, the daily price features are expressed as the simple returns:
\begin{equation}
DP^{sc}_t = \left[ \frac{O^{sc}_{t}}{C^{sc}_{t-1}} - 1, \frac{H^{sc}_{t}}{C^{sc}_{t-1}} - 1, \frac{L^{sc}_{t}}{C^{sc}_{t-1}} - 1, \frac{C^{sc}_{t}}{C^{sc}_{t-1}} - 1 \right ]
\label{eq:price_mode_vec}
\end{equation}

The sequence of historical corpus headlines $N^{sc}_t$ is expressed as
\begin{equation}
N^{sc}_t = \left [n^{sc}_{t-T}, n^{sc}_{t-T+1}, \cdots, n^{sc}_{t} \right]
\end{equation}
where $n^{sc}_{t^{\prime}}$ is a set containing all headlines that influence the market on a given day $t^{\prime}$.

Aiming to align prices and news modes, we consider the \textit{explicit alignment} method discussed in \ref{subsec:stock_headlines}. That is, $n^{sc}_{t^{\prime}}$ contains all stock headlines before the market opens ($\texttt{before market}_{t}$), during the trading hours \newline ($\texttt{during market}_{t}$), and previous day after-markets \newline ($\texttt{after market}_{t-1}$).

%After the corpus historical headlines $N_t$ are aligned we convert each sentence to a vector of integers were each element is a key to the respective pre-trained word embedding matrix. This process is described as follow: First, we tokenize each sentence and extract the corpus vocabulary size $\lvert V \rvert$. Second, we build the embedding matrix $E_w \in \mathbb{R}^{\lvert V \rvert \times d_w}$, where each row entry is a word embedding with $d_w$ dimensions. Finally, we assign each sentence word to an integer that points to its respective embedding.

As a text preprocessing step, we tokenize the headlines and convert each word to an integer that refers to its respective pre-trained word embedding. This process is described as follows: First, for all stocks of our corpus we tokenize each headline and extract the corpus vocabulary set $V$. We then build the embedding matrix $E_w \in \mathbb{R}^{\lvert V \rvert \times d_w}$, where each row is a word embedding vector $d_w$ dimensions. Words that do not have a corresponding embedding, i.e. out of vocabulary words, are skipped. 

%After the corpus historical headlines were aligned, we convert each headline $n_t$ to a matrix $l_s \times d_w$, where $l_s$ is the maximum length of a headline in the corpus. To this aim, we tokenize each sentence and convert each token to its word embedding representation (tokens without a corresponding word embedding are assigned to zero).

Finally, the input sample of the text mode is a tensor of integers with $T \times l_n \times l_s$ dimensions, where $l_n$ is the maximum number of news occurring in a given day and $l_s$ is the maximum length of a corpus sentence. Regarding the price mode, we have a $T \times 4$ tensor of floating numbers.

\subsection{Global features and stock embedding}
\label{sub:global_model}
Given the price and news histories for each stock $sc$ we could directly learn \textit{one model per stock}. However, this approach suffers from two main drawbacks. First, the market activity of one specific stock is expected to impact other stocks, which is a widely accepted pattern named ``spillover effect''. Second, since our price data is sampled on a daily basis, we would train the stock model relying on a small number of samples. One possible solution to model the commonality among stocks would be \textit{feature enrichment}. For example, when modeling a given stock $X$ we would enrich its news and price features by concatenating features from stock $Y$ and $Z$. Although the feature enrichment is able to model the effect of other stocks, it still would consider only one sample per day.

In this work, we propose a method that learns an \textit{global model}.

The global model is implemented using the following methods:

\begin{itemize}

\item \textbf{Multi-Stock batch samples}:
Since our models are trained using Stochastic Gradient Descent, we propose at each mini-batch iteration to sample from a batch set containing any stock of our stocks universe. As a consequence, the mapping between volatility and multimodal data is now able to learn common explanatory factors among stocks. Moreover, adopting this approach increases the total number of training samples, which is now the sum of the number of samples per stock.

\item \textbf{Stock Embedding}: Utilizing the Multi-Stock batch samples above, we tackle the problem of modeling commonality among stocks. However, it is reasonable to assume that stocks have part of its dynamic driven by idiosyncratic factors. Nevertheless, we could aggregate stocks per sector or rely on any measure of similarity among stocks. In order to incorporate information specific to each stock, we propose to equip our model with a ``stock embedding'' mode that is learnt jointly with price and news modes. That is to say, we leave the task of distinguishing the specific dynamic of each stock to be learnt by the neural network. Specifically, this stock embedding is modeled using a discrete encoding as input, i.e. $\mathcal{I}^{sc}_t$ is a vector with size equal to the number of stocks of the stocks universe and has element 1 for the i-th coordinate and 0 elsewhere, thus, indicating the stock of each sample.

\end{itemize}

Formally, we can express the one model per stock approach as the mapping
\begin{equation}
\begin{split}
\sigma^{sc}_{t+1} = f^{sc} ( DN^{sc}_{t-T}, DN^{sc}_{t-T+1}, \cdots, DN^{sc}_t ; \\
DP^{sc}_{t-T}, DP^{sc}_{t-T+1}, \cdots, DP^{sc}_t )
\end{split}
\end{equation}
where $DN^{sc}_{t^{\prime}}$ is a fixed-vector representing all news released on a given day for the stock $sc$\footnote{It will become clear in the next section how this news representation is modelled.} and $DP^{sc}_{t^{\prime}}$ is defined in \ref{eq:price_mode_vec}. 

The global model attempts to learn a single mapping $f$ that at each mini-batch iteration randomly aggregates samples across all the universe of stocks, rather than one mapping $f^{sc}$ per stock. The global model is expressed as
\begin{equation}
\begin{split}
\sigma^{sc}_{t+1} = f ( DN^{sc}_{t-T}, DN^{sc}_{t-T+1}, \cdots, DN^{sc}_t ; \\
DP^{sc}_{t-T}, DP^{sc}_{t-T+1}, \cdots, DP^{sc}_t ; \\
\mathcal{I}^{sc}_t)
\end{split}
\end{equation}

In the next section, we describe our hierarchical neural model and how the news, price and stock embedding are fused into a joint representation.
 
\subsection{Our multimodal hierarchical network}
\label{sub:HAN}
\begin{figure*}[ht]
\centering
\includegraphics[width=0.85\textwidth]{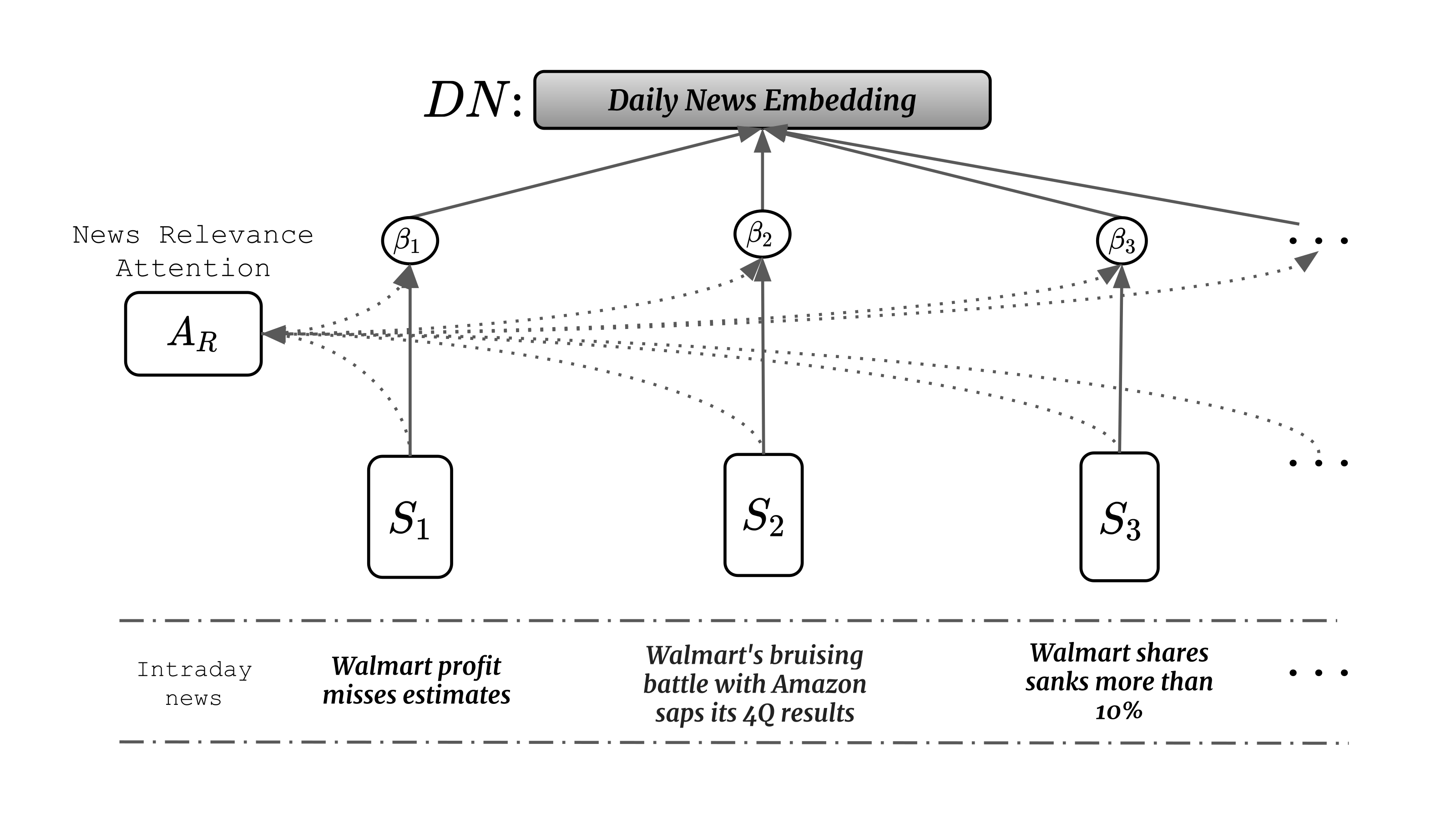}
\captionsetup{width=0.85\textwidth}
\caption{\textbf{Daily news relevance attention}. The figure illustrates a day where three news were released for the Walmart company. After the headlines are encoded into a fixed-size representation $S$, the daily \textit{news relevance attention} $A_R$ converts all sentences into single vector representation of all Daily News $DN$ by attending each headline based on its content.}
\label{fig:DN_encoder}
\end{figure*}

\begin{figure*}[ht]
\centering
\includegraphics[width=0.85\textwidth]{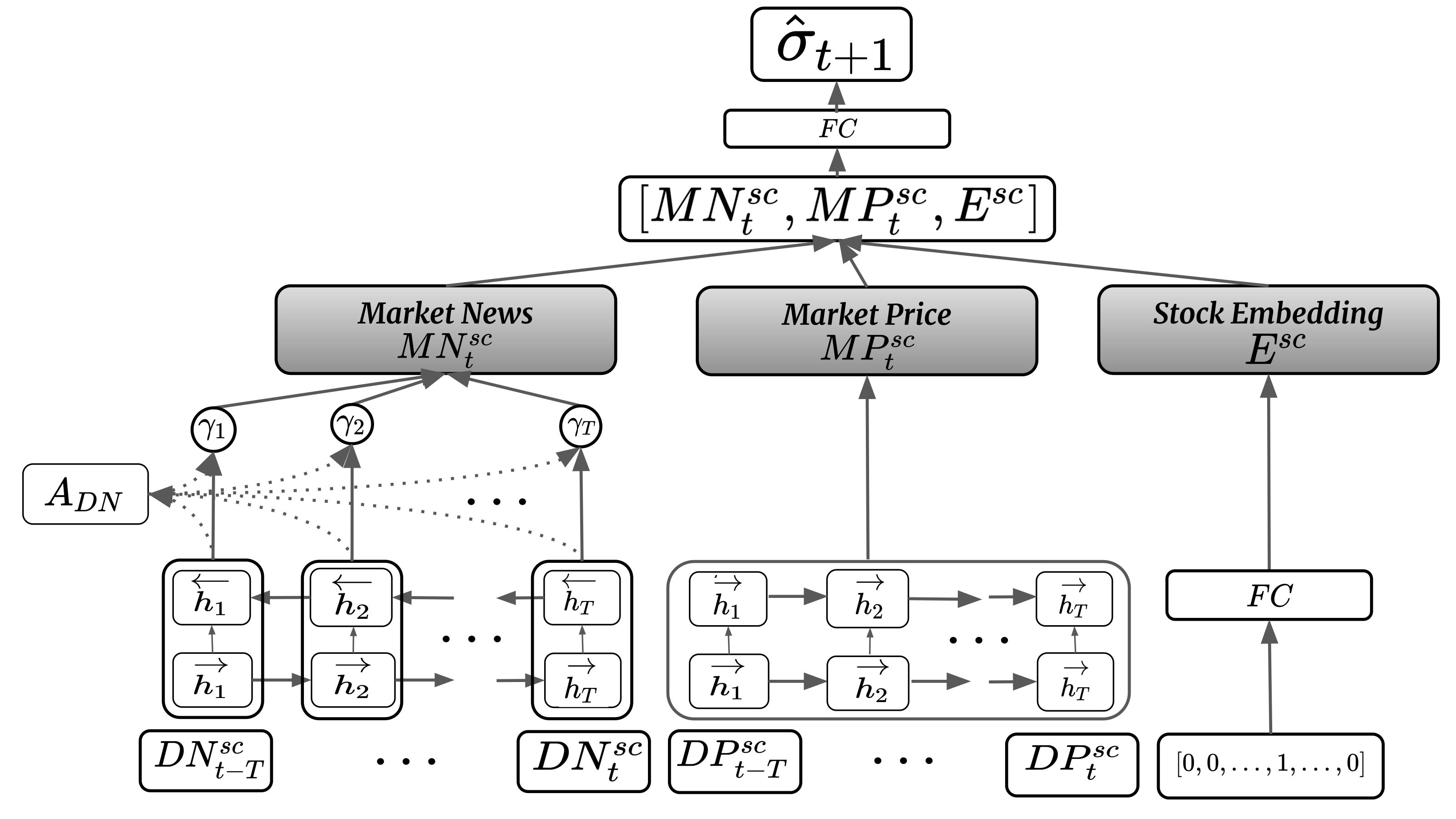}
\captionsetup{width=0.85\textwidth}
\caption{\textbf{Hierarchical Neural Network architecture.}}
\label{fig:nn_time_series_arch}
\end{figure*}

In broad terms, our hierarchical neural architecture is described as follows. First, each headline released on a given day $t$ is encoded into a fixed-size vector $S_t$ using a sentence encoder\footnote{The headline encoding $S_t$ is learnt end-to-end from the headline word embeddings or transfered from the TL tasks as fixed features.}. We then apply our daily New Relevance Attention (NRA) mechanism that attends each news based on its content and converts a variable size of news released on a given day into a single vector denoted by Daily News ($DN$). We note that this representation take account of the overall effect of all news released on a given day. This process is illustrated in \ref{fig:DN_encoder}. We now are in a position to consider the temporal effect of the past $T$ days of market news and price features. \ref{fig:nn_time_series_arch} illustrates the neural network architecture from the temporal sequence to the final volatility prediction. For each stock code $sc$ the temporal encoding for news is denoted by Market News $MN^{sc}_t$ and for the price by Market Price $MP^{sc}_t$ and are a function of the past $T$ Daily News representations ${\{DN^{sc}_{t-T}, \cdots, DN^{sc}_t \}}$ (\textbf{Text mode}) and Daily Prices features ${\{DP^{sc}_{t-T}, \cdots, DP^{sc}_t \}}$ (\textbf{Price mode}), where each Daily Price $DP^{sc}_{t^{\prime}}$ feature is given by \ref{eq:price_mode_vec} and the $DN^{sc}_{t^{\prime}}$ representation is calculated using \textbf{Daily New Relevance Attention}. After the temporal effects of $T$ past days of market activity were already encoded into the Market News $MN^{sc}_t$ and Market Price $MP^{sc}_t$, we concatenate feature-wise $MN^{sc}_t$, $MP_t$ and the \textbf{Stock embedding} $E^{sc}$. The stock embedding $E^{sc}$ represents the stock code of the sample on a given day $t$. Finally, we have a Fully Connected (FC) layer that learns the \textbf{Joint Representation} of all modes. This fixed-sized joint representation is fed into a FC layer with linear activation that predicts the next day volatility $\hat{\sigma}_{t+1}$.

Below, we detail, for each mode separately, the layers of our hierarchical model. 

-- \textbf{Text mode}
\begin{enumerate}[1.]

\item \textbf {Word Embedding Retrieval} \newline
Standard embedding layer with no trainable parameters. It receives a vector of word indices as input and returns a matrix of word embeddings.

\item \textbf{News Encoder} \newline
This layer encodes all news on a given day and outputs a set news embeddings $\{S^{1}_t, \cdots, S^{l_n}_t \}$. Each encoded sentence has dimension $d_S$, which is a hyperparameter of our model. This layer constitutes a key component of our neural architectures and, as such, we evaluate our models considering sentence encoders trained end-to-end, using the BiLSTM attention (\ref{subsec:bilstm_inner_att}) and BiLSTM max-pooling (\ref{subsec:bilstm_max_pool}) architectures, and also transferred from the RCV1 and SNLI as fixed features. 

\item \textbf{Daily news relevance attention} \newline
Our proposed \textbf{news relevance} attention mechanism for all news released on a given day. The attention mechanism is introduced to tackle information overload. It was designed to ``filter out'' redundant or misleading news and focus on the relevant ones based solely on the news content. Formally, the layer outputs a Daily News (DN) embedding $DN^{sc}_t = \sum_{i=1}^{l_n} \beta_i S^{sc^{i}}_t$, which is a linear combination of all encoded news on a given day $t$. This news-level attention uses the same equations as in \ref{eq:att}, but with trainable weights $\{W_{R}, b_{R}, v_{R}\}$, i.e. the weights are segregated from the sentence encoder. \ref{fig:DN_encoder}, illustrates our relevance attention. Note that this layer was deliberately developed to be invariant to headlines permutation, as is the case with the linear combination formula above. The reason is that our price data is sampled daily and, as a consequence, we are not able to discriminate the market reaction for each intraday news.     

\item \textbf{News Temporal Context} \newline
\label{itm:news_tc_layer}
Sequence layer with daily news embeddings $DN^{sc}_t$ as time steps. This layer aims to learn the temporal context of news, i.e. the relationship between the news at day $t$ and the $T$ past days. It receives as input a chronologically ordered sequence of $T$ past Daily News embeddings ${\{DN^{sc}_{t-T}, \cdots, DN^{sc}_t \}}$ and outputs the news mode encoding Market News $MN^{sc}_t \in d_{MN}$. The sequence with $T$ time steps is encoded using a BiLSTM attention. The layer was designed to capture the temporal order that news are released and the current \textbf{news novelty}. i.e. news that were repeated in the past can be ``forgotten'' based on the modulating gates of the LSTM network. 

%Note that this layer is designed to capture two crucial memory components. First, it is expected that the volatility prediction is sensitive to the order that the news were released, in the same manner that the sentence encoder depends on the order of the words. We call this effect the \textbf{news temporal context}. Second, the LSTM is equipped with a memory state able to learn  \textbf{news novelty}, i.e. news that were repeated in the past can be ``forgotten'' based on the modulating gates of the LSTM network.
\end{enumerate}

-- \textbf{Price mode}
\begin{enumerate}[5.] % start counting from the previous one
\item \textbf{Price Encoder} \newline
Sequence layer analogous to \textbf{News Temporal Context}, but for the price mode. The input is the ordered sequence Daily Prices ${\{DP^{sc}_{t-T}, \cdots, DP^{sc}_t \}}$ of size $T$, where each element the price feature defined in \ref{eq:price_mode_vec}. Particularly, the architecture consists of two stacked LSTM's.  The first one outputs for each price feature time step a hidden vector that takes the temporal context into account. Then these hidden vectors are again passed to a second independent LSTM. The layer outputs the price mode encoding Market Price $MP^{sc}_t \in d_{MP}$. This encoding is the last hidden vector of the second LSTM Market.
\end{enumerate}

-- \textbf{Stock embedding}
\begin{enumerate}[6.] % start counting from the previous one
\item \textbf{Stock Encoder} \newline
Stock dense representation. The layer receives the discrete encoding $\mathcal{I}^{sc}_t$ indicating the sample stock code pass through a FC layer and outputs a stock embedding $E_{sc}$. 
\end{enumerate}

-- \textbf{Joint Representation}
\begin{enumerate}[7.] % start counting from the previous one
\item \textbf{Merging} \newline
Feature-wise News, Price, and Stock modes concatenation. No trainable parameters. 
\item \textbf{Joint Representation Encoder} \newline
FC layer of size $d_{JR}$. 
\end{enumerate}

\subsection{Multimodal learning with missing modes}

During the training we feed into our neural model the price, news, and stock indicator data. The price and stock indicator modes data occur in all days. However, at the individual stock level we can have days that the company is not covered by the media. This feature imposes challenges to our multimodal training since neural networks are not able to handle missing modes without special intervention. A straightforward solution would be to consider only days with news released, disregarding the remaining samples. However, this approach has two main drawbacks. First, the ``missing news'' do not happen at random, or are attributed to measurement failure as is, for example, the case of multimodal tasks using mechanical sensors data. Conversely, as highlighted in \cite{Chan2003,Boudoukh2013} the same price behaviour results in distinct market reactions when accompanied or not by news\footnote{Experimental results \cite{Chan2003,Boudoukh2013} demonstrate that large price dislocations in the absence of news tend revert and continue the movement (momentum) when driven by news.}. In other words, specifically to financial forecasting problems the absence or existence of news are highly informative.  

Some methods were proposed in the multimodal literature to effectively treat informative missing modes or ``informative missingness'', which is a characteristic refereed in the literature as \textit{learning with missing modalities} \cite{Baltrusaitis2017}.
In this work, we directly model the news missingness as a feature of our text model temporal sequence by using the method initially proposed in \cite{Lipton2016,LiptonClinical2016} for clinical data with missing measurements and applied in the context of financial forecasting in \cite{Alberg2017a}. Specifically, we implement the \textit{Zeros \& Imputation} (ZI) method \cite{LiptonClinical2016} in order to jointly learn the price mode and news relationship across all days of market activity.

The ZI implementation is described as follows: Before the daily news sequence is processed by the text temporal layer (described in \ref{itm:news_tc_layer}) we input a 0 vector for all time steps with missing news and leave the news encoding unchanged otherwise. This step is called zero imputation. In addition, we concatenate feature-wise an indicator vector with value 1 for all vectors with zero imputation and 0 for the days with news.

As described in \cite{Alberg2017a}, the ZI method endow a temporal sequence model with the ability to learn different representations depending on the news history and its relative time  position. Moreover, it allows our model to predict the volatility for all days of our time series and, at the same time, take into account the current and past news informative missingness. Furthermore, the learnt positional news encoding works differently than a typical ``masking'', where days without news are not passed through the LSTM cell. Masking the time steps would be losing information about the presence or absence of news concomitant with prices.

\section{Experimental results and discussions}

We aim to evaluate our hierarchical neural model in the light of three main aspects. First, we asses the importance of the different sentence encoders to our end-to-end models and how it compares to transferring the sentence encoder from our two auxiliary TL tasks. Second, we ablate our proposed news relevance attention (NRA) component to evaluate its importance. Finally, we consider a model that takes into consideration only the price mode (unimodal), i.e. ignoring any architecture related to the text mode.

Before we define the baselines to asses the three aspects described above, we review in the next section the scores of the trained TL tasks.

\subsection{Auxiliary transfer learning tasks}
This section reports the performance of the auxiliary TL tasks considered in this work. Our ultimate goal is to indicate that our scores are in line with previous works%, thus, the trained TL sentence encoders provide reliable sentence representations.

All the architectures presented in \ref{sec:transfer_learning} are trained for a maximum of 50 epochs using mini-batch SGD with Adam optimizer \cite{Kingma2015}. Moreover, at the end of each epoch, we evaluate the validation scores, which are accuracy (Stanfor SNLI dataset) and F1 (RCV1 dataset), and save the weights with the best values. Aiming to seeped up training, we implement \textit{early stopping} with patience set to 8 epochs. That is, if the validation scores do not improve for more than 10 epochs we halt the training. Finally, we use Glove pre-trained word embeddings \cite{Pennington2014} as fixed features. 

\ref{tbl:tl_evaluation} compares our test scores with state-of-the-art (SOTA) results reported in previous works. We can see that our scores for the SNLI task are very close to state-of-the-art\footnote{Models were trained using a concatenation layer and Bidirectional LSTM with 512 and 1024 units, respectively}.

Regarding the RCV1 dataset, our results consider only the headline content for training, while the refereed works consider both the news headline and message body. The reason for training using only the headlines is that both tasks are learnt with the sole purpose of transferring the sentence encoders to our main volatility prediction task, whose textual input is restricted to headlines. 

\begin{table*}[ht]
\centering
\resizebox{\textwidth}{!}{\begin{minipage}{1.15\textwidth}
\begin{tabularx}{1.0\textwidth}{cXc}
\toprule
% \rotatebox[origin=c]{90}{\textit{SNLI}}
Dataset & Sentence Encoder & Score \\
\midrule
\multirow{6}{*}{\rotatebox[origin=c]{90}{\textit{SNLI}}} & LSTM original paper (\cite{Bowman2015}) & 0.806 \\
& BiLSTM over Mean Pooling (\cite{Liu2016}) & 0.833 \\
& BiLSTM attention (Att) with multiple views and factored fusion layer (\cite{Lin2017}) & 0.844 \\
& BiLSTM max-pooling (MP) with sentence embedding size 4096 (\cite{Conneau2017}) & 0.845 \\
& Our BiLSTM Att with sentence embedding size 2048 &  \underline{0.838} \\
& Our BiLSTM MP with sentence embedding size 2048 & \underline{0.841} \\
\midrule
\multirow{5}{*}{\rotatebox[origin=c]{90}{\textit{RCV1}}} & $k$-NN\textsuperscript{$\dagger$} (\cite{Lewis2004}) & 0.765 \\
& Best Support Vector Machine (SVM)\textsuperscript{$\dagger$} (\cite{Lewis2004}) & 0.816 \\
& bow-CNN\textsuperscript{$\dagger$} (\cite{Johnson2015}) & 0.840 \\
%\cmidrule{lr}{2-3}
& Our BiLSTM Att with sentence embedding size 2048 (headlines only) & \underline{0.809} \\
& Our BiLSTM MP with sentence embedding size 2048 (headlines only) & \underline{0.811} \\
\bottomrule
\end{tabularx}
\captionsetup{width=1.0\textwidth}
\caption{\textbf{TL auxiliary tasks -- Sentence Encoders comparison.} Test scores are accuracy and F1 scores for the SNLI \ref{subsec:stanford_nli} and RCV1 \ref{subsec:reuters_rcv1} datasets, respectively. $\dagger$ indicates model trained with both headlines and body content and using the original 103 classes of the RCV1 dataset, rather than our models that are trained using headlines only and a total of 55 classes (see \ref{subsec:reuters_rcv1} for a complete description). As a consequence, the reported benchmarks for the RCV1 dataset are not directly comparable and where reported for the sake of a better benchmark.}
\label{tbl:tl_evaluation}
\end{minipage}}
\end{table*}

\subsection{Training setup}
During the training of our hierarchical neural model described in \ref{sub:HAN} we took special care to guard against overfitting. To this aim, we completely separate 2016 and 2017 as the test set and report our results on this ``unseen'' set. The remaining data is further split into training (2007 to 2013) and validation (2014 to 2015). The model convergence during training is monitored in the validation set. We monitor the validation score of our model at the end of each epoch and store the network weights if the validation scores improves between two consecutive epochs. Additionally, we use mini-batch SGD with Adam optimizer and early stopping with patience set to eight epochs. The hyperparameter tunning is performed using grid search.

All training is performed using the proposed global model approach described in \ref{sub:global_model}, which learns a model that takes into account the features of all the 40 stocks of our corpus. Using this approach our training set has a total of 97,903 samples. Moreover, during the SGD mini-batch sampling the past $T$ days of price and news history tensors and each stock sample stock indicator are randomly selected from the set of all 40 stocks.   

\subsection{Stocks universe result}   
In order to evaluate the contributions of each component of our neural model described in \ref{sub:HAN} and the effect of using textual data to predict the volatility, we report our results using the following baselines\footnote{Minus sign means to remove (ablate) the neural network component while plus means to include the component.}:

\begin{enumerate}[1.]

\item \textbf{- News (unimodal price only)}: This baseline completely ablates (i.e. removes) any architecture related to the news mode, considering only the price encoding and the stock embedding components. Using this ablation we aim to evaluate the influence of news to the volatility prediction problem.

\item \textbf{+ News (End-to-end Sentence Encoders) - NRA}: This baseline ablates our proposed new relevance attention (NRA) component, and instead, makes use of the same \textit{Daily Averaging} method in \cite{Ding2015,Pinheiro2017}, where all fixed-sized headline representations on a given day are averaged without taking into account the relevance of each news. We evaluate this baseline for both BiLSTM attention (Att) and BiLSTM max-pooling (MP) sentence encoders. Here, our goal is to asses the true contribution of our NRA component in the case SOTA sentence encoders are taken into account.

\item \textbf{+ News (End-to-End W-L Att Sentence Encoder) + NRA}: The Word-Level Attention (W-L Att) sentence encoder implements an attention mechanism directly on top of word embeddings, and, as such, does not consider the order of words in a sentence. This baseline complements the previous one, i.e. it evaluates the influence of the sentence encoder when our full specification is considered.  

\item \textbf{+ News (TL Sentence Encoders) + NRA}: Makes use of sentence encoders of our two auxiliary TL tasks as fixed features. This baseline aims to address the following questions, namely: What dataset and models are more suitable to transfer to our specific volatility forecasting problem; How End-to-End models, which are trained on top of word embeddings, perform compared to sentence encoders transferred from other tasks.

\end{enumerate}
  
\ref{tbl:comparative_all_sectors} summarizes the test scores for the ablations discussed above. Our best model is the + News (BiLSTM Att) + NRA, which is trained end-to-end and uses our full architecture. The second best model, i.e. + News  (BiLSTM MP) + NRA, ranks slightly lower and only differs form the best model in terms of the sentence encoder. The former sentence encoder uses an attention layer (\ref{subsec:bilstm_inner_att}) and the the last a max-pooling layer (\ref{subsec:bilstm_max_pool}), where both layers are placed on top of the LSTM hidden states of each word. 

Importantly, our experiments show that using news and price (multimodal) to predict the volatility improves the scores by 11\% (MSE) and 9\% (MAE) when compared with the – News (price only unimodal) model that considers only price features as explanatory variables. 

When comparing the performance of End-to-End models and the TL auxiliary tasks the following can be observed: The end-to-end models trained with the two SOTA sentence encoders perform better than transferring sentence encoder from both auxiliary tasks. However, our experiments show that the same does not hold for models trained end-to-end relying on the simpler WL-Att sentence encoder, which ignores the order of words in a sentence. In other words, considering the appropriate TL task, it is preferable to transfer a SOTA sentence encoder trained on a larger dataset than learning a less robust sentence encoder in an end-to-end fashion. Moreover, initially, we thought that being the RCV1 a financial domain corpus it would demonstrate a superior performance when compared to the SNLI dataset. Still, the SNLI transfers better than RCV1. We hypothesize that the text categorization task (RCV1 dataset) is not able to capture complex sentence structures at the same level required to perform natural language inference. Particularly to the volatility forecasting problem, our TL results corroborates the same findings in \cite{Conneau2017}, where it was shown that SNLI dataset attains the best sentence encoding for a broad range of pure NLP tasks, including, among other, text categorization and sentiment analysis.

Significantly, experimental results in \ref{tbl:comparative_all_sectors} clearly demonstrate that our proposed \textit{news relevance attention} (NRA) outperforms the News Averaging method proposed in previous studies \cite{Ding2015,Pinheiro2017}. Even when evaluating our NRA component in conjunction with the more elementary W-L Att sentence encoder it surpass the results of sophisticated sentence encoder using a News Averaging approach. In other words, our results strongly points to the advantage of discriminating noisy from impacting news and the effectiveness of learning to attend the most relevant news.

\begin{table*}[b]
\centering
\resizebox{\textwidth}{!}{\begin{minipage}{1.10\textwidth}
%\begin{tabularx}{\textwidth}{C{0.02}L{0.50}C{0.12}C{0.12}C{0.12}C{0.12}}
\begin{tabularx}{\textwidth}{Xcc}
\toprule 
Model & MSE & MAE \\
\midrule
\multicolumn{3}{c}{\textbf{All stocks}} \\
\midrule
- News (price only unimodal)\textsuperscript{$\dagger$} & 2.140E-05 & 3.093E-03 \\
+ News (BiLSTM Att) - \textit{news relevance attention} (NRA) & 2.078E-03 & 3.037E-03 \\
+ News (BiLSTM MP) - NRA & 2.077E-03 & 3.031E-03 \\
+ News (TL Reuters RCV1 BiLSTM MP) + NRA & 2.037E-03 & 3.020E-03 \\
+ News (TL Reuters RCV1 BiLSTM Att) + NRA  & 2.023E-03 & 3.011E-03 \\
+ News (W-L Att)\textsuperscript{$\dagger\dagger$} + NRA  & 2.006E-03 & 2.947E-03 \\
+ News (TL SNLI BiLSTM Att) + NRA  & 1.986E-03 & 2.926E-03 \\
+ News (TL SNLI BiLSTM MP) + NRA  & 1.974E-03 & 2.918E-03 \\
+ News (BiLSTM MP) + NRA & 1.904E-03 & 2.851E-03 \\
\textbf{+ News (BiLSTM Att) + NRA} & \textbf{1.898E-03} & \textbf{2.823E-03} \\
\bottomrule
\end{tabularx}
\captionsetup{width=\textwidth}
\caption{\textbf{Model architecture ablations and sentence encoders comparisons}. The minus sign means that the component of our network architecture described in \ref{sub:HAN} was ablated (i.e. removed) and the plus sign that it is added. The second and third row report results replacing the \textit{news relevance attention} (NRA) with a News Averaging component as in \cite{Ding2015,Pinheiro2017}. $\dagger$ indicates our model was trained using only the price mode. $\dagger\dagger$ highlights that the sentence encoder Word-Level Attention (W-L Attention) does not take into consideration the headline words order. Best result in bold.}
\label{tbl:comparative_all_sectors}
\end{minipage}}
\end{table*}

Having analyzed our best model, we now turn to its comparative performance with respect to the widely regarded GARCH(1,1) model described in \ref{sec:GARCH}.

We asses our model performance relative to GARCH(1,1) using standard loss metrics (MSE and MAE) and the regression-based accuracy specified in \ref{eq:regression_loss} and measured in terms of the coefficient of determination $R^2$. In addition, we evaluate our model across two different volatility proxies: Garman-Klass ($\widehat{\sigma_{GK}}$) (\ref{eq:vol_gk}) and Parkinson ($\widehat{\sigma_{PK}}$) (\ref{eq:vol_pk}). We note that, as reviewed in \ref{sub:evalution}, these two volatility proxies are statically efficient and proper estimators of the next day volatility.

\ref{tbl:garch_all_sectors} reports the comparative performance among our best Price + News model (+ News BiLSTM (MP) + NRA), our Price only (unimodal) model and GARCH(1,1). The results clearly demonstrate the superiority of our model, being more accurate than GRACH for both volatility proxies. We note that evaluating the GARCH(1,1) model relying on standard MSE and MAE error metrics should be taken with a grain of salt. \cite{Andersen1998} provides the background theory and arguments supporting $R^2$ as the metric of choice to evaluate the predictive power of a volatility model. In any case, the outperformance or our model with respect to GARCH(1,1) permeates all three metrics, name $R^2$, $MSE$ and $MAE$.

\begin{table*}[ht]
\centering
\begin{tabularx}{1.0\textwidth}{Xcccc}
\toprule 
\multirow{2}{*}{Model} & Vol & \multirow{2}{*}{$R^2$} & \multirow{2}{*}{MSE} & \multirow{2}{*}{MAE} \\
& Estimator & & & \\
\midrule
\multicolumn{5}{c}{\textbf{All Stocks}} \\
\midrule
\multirow{2}{*}{\textit{GARCH(1,1)}} & $\widehat{\sigma_{GK}}$ & 0.357 & 2.46E-05 & 3.16E-03 \\
& $\widehat{\sigma_{PK}}$ & 0.329 & 2.57E-05 & 3.20E-03 \\
\midrule
\multirow{2}{*}{\textit{Our Model: Price (Unimodal)}} & $\widehat{\sigma_{GK}}$ & 0.384 & 2.14E-05 & 3.09E-03 \\
& $\widehat{\sigma_{PK}}$ & 0.350 & 2.36E-05 & 3.29E-03 \\
\midrule
\multirow{2}{*}{\textbf{\textit{Our Model: Price + News}}} & $\widehat{\sigma_{GK}}$ & \textbf{0.455}	& \textbf{1.90E-05}	& \textbf{2.82E-03} \\
& $\widehat{\sigma_{PK}}$ & \textbf{0.410} & \textbf{2.09E-05} & \textbf{2.98E-03} \\
\bottomrule
\end{tabularx}
\captionsetup{width=1.0\textwidth}
\caption{\textbf{Our volatility model performance compared with GARCH(1,1)}. Best performance in bold. Our model has superior performance across the three evaluation metrics and taking into consideration the state-of-the-art volatility proxies, namely Garman-Klass ($\widehat{\sigma_{PK}}$) and Parkinson ($\widehat{\sigma_{PK}}$).}
\label{tbl:garch_all_sectors}
\end{table*}

\subsection{Sector-level results}
Company sectors are expected to have different risk levels, in the sense that each sector is driven by different types of news and economic cycles. Moreover, by performing a sector-level analysis we were initially interested in understanding if the outperformance of our model with respect to GARCH(1,1) was the result of a learning bias to a given sector or if, as turned out to be the case, the superior performance of our model spreads across a diversified portfolio of sectors.  

In order to evaluate the performance per sector, we first separate the constituents stocks for each sector in \ref{tbl:stock_universe}. Then, we calculate the same metrics discussed in the previous section for each sector individually. 

\ref{tbl:garch_each_sector} reports our experimental results segregated by sector. We observe that the GRACH model accuracy, measured using the $R^2$ score, has a high degree of variability among sectors. For example, the accuracy ranges from 0.15 to 0.44 for the HealthCare and Energy sector, respectively. This high degree of variability is in agreement with previous results reported in \cite{Rekabsaz2017}, but in the context of long-term (quarterly) volatility predictions. Although the GARCH(1,1) accuracy is sector-dependent, without any exception, our model using price and news as input clearly outperforms GRACH sector-wise. This fact allow us to draw the following conclusions:
\begin{itemize}
\item Our model outperformance is persistent across sectors, i.e. the characteristics of the results reported in \ref{tbl:garch_all_sectors} permeates all sectors, rather than being composed of a mix of outperforming and underperforming sector contributions. This fact provides a strong evidence that our model is more accurate than GARCH(1,1).
\item The proposed Global model approach discussed in \ref{sub:global_model} is able to generalize well, i.e. the patterns learnt are not biased to a given sector or stock.
\end{itemize}

\begin{table*}[b]
\centering
\resizebox{0.55\textheight}{!}{\begin{minipage}{\textwidth}
%\begin{tabularx}{\textwidth}{}
%\toprule 
\begin{tabularx}{\textwidth}{Xcccc}
\toprule 
\multirow{2}{*}{Model} & Vol & \multirow{2}{*}{$R^2$} & \multirow{2}{*}{MSE} & \multirow{2}{*}{MAE} \\
& Estimator & & & \\
\midrule
% Consumer Staples
\multicolumn{5}{c}{\textbf{Consumer Staples}} \rule{0pt}{5ex} \\
\midrule
\multirow{2}{*}{\textit{GARCH(1,1)}} & $\widehat{\sigma_{GK}}$ & 0.173 & 2.01E-05 & 2.63E-03 \\
& $\widehat{\sigma_{PK}}$ & 0.155 & 2.08E-05 & 2.70E-03 \\
\midrule
\multirow{2}{*}{\textit{Our Model: Price (Unimodal)}} & $\widehat{\sigma_{GK}}$ & 0.194 & 1.93E-05 & 2.67E-03 \\
& $\widehat{\sigma_{PK}}$ & 0.176 & 2.04E-05 & 2.82E-03 \\
\midrule
\multirow{2}{*}{\textbf{\textit{Our Model: Price + News}}} & $\widehat{\sigma_{GK}}$ & \textbf{0.224} & \textbf{1.80E-05} & \textbf{2.48E-03} \\
& $\widehat{\sigma_{PK}}$ & \textbf{0.201} & \textbf{1.90E-05} & \textbf{2.61E-03} \\
% HealthCare
\multicolumn{5}{c}{\textbf{HealthCare}} \rule{0pt}{5ex} \\
\midrule
\multirow{2}{*}{\textit{GARCH(1,1)}} & $\widehat{\sigma_{GK}}$ & 0.150 & 2.20E-05 & 3.05E-03 \\
& $\widehat{\sigma_{PK}}$ & 0.138 & 2.33E-05 & 3.09E-03 \\
\midrule
\multirow{2}{*}{\textit{Our Model: Price (Unimodal)}} & $\widehat{\sigma_{GK}}$ & 0.186 & 2.01E-05 & 3.01E-03 \\
& $\widehat{\sigma_{PK}}$ & 0.164 & 2.24E-05 & 3.21E-03 \\
\midrule
\multirow{2}{*}{\textbf{\textit{Our Model: Price + News}}} & $\widehat{\sigma_{GK}}$ & \textbf{0.258} & \textbf{1.76E-05} & \textbf{2.74E-03} \\
& $\widehat{\sigma_{PK}}$ & \textbf{0.225} & \textbf{1.96E-05} & \textbf{2.90E-03} \\
% Financials
\multicolumn{5}{c}{\textbf{Financials}} \rule{0pt}{5ex} \\
\midrule
\multirow{2}{*}{\textit{GARCH(1,1)}} & $\widehat{\sigma_{GK}}$ & 0.274 & 2.02E-05 & 3.14E-03 \\
& $\widehat{\sigma_{PK}}$ & 0.250 & 2.17E-05 & 3.18E-03 \\
\midrule
\multirow{2}{*}{\textit{Our Model: Price (Unimodal)}} & $\widehat{\sigma_{GK}}$ & 0.326 & 1.77E-05 & 3.10E-03 \\
& $\widehat{\sigma_{PK}}$ & 0.290 & 2.03E-05 & 3.32E-03 \\
\midrule
\multirow{2}{*}{\textbf{\textit{Our Model: Price + News}}} & $\widehat{\sigma_{GK}}$ & \textbf{0.373}	& \textbf{1.65E-05} & \textbf{2.84E-03} \\
& $\widehat{\sigma_{PK}}$ & \textbf{0.332} & \textbf{1.86E-05} & \textbf{3.00E-03} \\
% Energy
\multicolumn{5}{c}{\textbf{Energy}} \rule{0pt}{5ex} \\
\midrule
\multirow{2}{*}{\textit{GARCH(1,1)}} & $\widehat{\sigma_{GK}}$ & 0.443 & 4.38E-05 & 4.24E-03 \\
& $\widehat{\sigma_{PK}}$ & 0.412 & 4.52E-05 & 4.27E-03 \\
\midrule
\multirow{2}{*}{\textit{Our Model: Price (Unimodal)}} & $\widehat{\sigma_{GK}}$ & 0.440 & 3.60E-05 & 4.13E-03 \\
& $\widehat{\sigma_{PK}}$ & 0.406 & 3.98E-05 & 4.34E-03 \\
\midrule
\multirow{2}{*}{\textbf{\textit{Our Model: Price + News}}} & $\widehat{\sigma_{GK}}$ & \textbf{0.538} & \textbf{3.04E-05} & \textbf{3.72E-03} \\
& $\widehat{\sigma_{PK}}$ & \textbf{0.495} & \textbf{3.38E-05} & \textbf{3.88E-03} \\
% Utilities
\multicolumn{5}{c}{\textbf{Utilities}} \rule{0pt}{5ex} \\
\midrule
\multirow{2}{*}{\textit{GARCH(1,1)}} & $\widehat{\sigma_{GK}}$ & 0.167 & 1.71E-05 & 2.75E-03 \\
& $\widehat{\sigma_{PK}}$ & 0.154 & 1.75E-05 & 2.77E-03 \\
\midrule
\multirow{2}{*}{\textit{Our Model: Price (Unimodal)}} & $\widehat{\sigma_{GK}}$ & 0.145 & 1.40E-05 & 2.56E-03 \\
& $\widehat{\sigma_{PK}}$ & 0.128 & 1.51E-05 & 2.75E-03 \\
\midrule
\multirow{2}{*}{\textbf{\textit{Our Model: Price + News}} }& $\widehat{\sigma_{GK}}$ & \textbf{0.225} & \textbf{1.24E-05} & \textbf{2.34E-03} \\
& $\widehat{\sigma_{PK}}$ & \textbf{0.193} & \textbf{1.34E-05} & \textbf{2.51E-03} \\
\bottomrule
\end{tabularx}
\captionsetup{width=1.0\textwidth}
\caption{\textbf{Sector-level performance comparison.}}
\label{tbl:garch_each_sector}
\end{minipage}}
\end{table*}

One of the limitations of our work is to rely on proxies for the volatility estimation. Although these proxies are handy if only open, high, low and close daily price data is available, having high frequency price data we could estimate the daily volatility using the sum of squared intraday returns to  measure the true daily latent volatility. For example, in evaluating the performance for the one-day-ahead GARCH(1,1) Yen/Dollar exchange rate \cite{Andersen1998} reports $R^2$ values of 0.237 and 0.392 using hourly and five minutes sampled intraday returns, respectively. However, we believe that utilizing intraday data would further improve our model performance. 

Since our experimental results demonstrate the key aspect of the \textit{news relevance attention} to model architecture we observe that intraday data would arguably ameliorate the learning process. Having intraday data would allow us to pair each individual news release with the instantaneous market price reaction. Using daily data we are losing part of this information by only measuring the aggregate effect of all news to the one-day-ahead prediction. 

\section{Conclusion}
We study the joint effect of stock news and prices on the daily volatility forecasting problem. To the best of our knowledge, this work is one of the first studies aiming to predict short-term (daily) rather than long-term (quarterly or yearly) volatility taking news and price as explanatory variables and using a comprehensive dataset of news headlines at the individual stock level. %Aiming to contribute to the research in the intersection between Finance and NLP, we make dataset, compiled at stock level and covering a diversified range of sectors, publicly available.

Our hierarchical end-to-end model benefits from state-of-the-art approaches to encode text information and to deal with two main challenges in correlating news with market reaction: news relevance and novelty. That is, to address the problem of how to attend the most important news based purely on its content (\textit{news relevance attention}) and to take into account the temporal information of past news (temporal context). Additionally, we propose a multi-stock mini-batch + stock embedding method suitable to model commonality among stocks.

The experimental results show that our multimodal approach outperforms the GARCH(1,1) volatility model, which is the most prevalent econometric model for daily volatility predictions. The outperformance being sector-wise and demonstrates the effectiveness of combining price and news for short-term volatility forecasting. The fact that we outperform GARCH(1,1) for all analyzed sectors confirms the robustness of our proposed architecture and evidences that our global model approach generalizes well.

We ablated (i.e. removed) different components of our neural architecture to assess its most relevant parts. To this aim, we replaced our proposed \textit{news relevance attention} layer, which aims to attend the most important news on a given day, with a simpler architecture proposed in the literature, which averages the daily news. We found that our attention layer improves the results. Additionally, we ablated all the architecture related to the news mode and found that news enhances the forecasting accuracy.

Finally, we evaluated different sentence encoders, including those transfered from other NLP tasks, and concluded that they achieve better performance as compared to a plain Word-level attention sentence encoder trained end-to-end. However, they do not beat state-of-the-art sentence encoders trained end-to-end.

In order to contribute to the literature of Universal Sentence Encoders, we  evaluated the performance of transferring sentence encoders from two different tasks to the volatility prediction problem. We showed that models trained on the Natural Language Inference (NLI) task are more suitable to forecasting problems than a financial domain dataset (Reuters RCV1). By analyzing different architectures, we showed that a BiLSTM with max-pooling for the SNLI dataset provides the best sentence encoder.

In the future, we plan to make use of intraday prices to better assess the predictive power of our proposed models. Additionally, we would further extend our analysis to other stock market sectors.

\section*{References}

\bibliography{mybibfile}

\begin{thebibliography}{10}
\expandafter\ifx\csname url\endcsname\relax
  \def\url#1{\texttt{#1}}\fi
\expandafter\ifx\csname urlprefix\endcsname\relax\def\urlprefix{URL }\fi
\expandafter\ifx\csname href\endcsname\relax
  \def\href#1#2{#2} \def\path#1{#1}\fi

\bibitem{Xing2018}
F.~Z. Xing, E.~Cambria, R.~E. Welsch,
  \href{http://link.springer.com/10.1007/s10462-017-9588-9}{{Natural language
  based financial forecasting: a survey}}, Artificial Intelligence Review
  50~(1) (2018) 49--73.
\newblock \href {http://dx.doi.org/10.1007/s10462-017-9588-9}
  {\path{doi:10.1007/s10462-017-9588-9}}.
\newline\urlprefix\url{http://link.springer.com/10.1007/s10462-017-9588-9}

\bibitem{Milgrom1982}
P.~Milgrom, N.~Stokey,
  \href{http://www.sciencedirect.com/science/article/pii/0022053182900461}{{Information,
  trade and common knowledge}}, Journal of Economic Theory.
\newline\urlprefix\url{http://www.sciencedirect.com/science/article/pii/0022053182900461}

\bibitem{Harris1993}
M.~Harris, A.~Raviv,
  \href{http://rfs.oxfordjournals.org/content/6/3/473.abstract}{{Differences of
  Opinion Make a Horse Race}}, Review of Financial Studies 6~(3) (1993)
  473--506.
\newblock \href {http://dx.doi.org/10.1093/rfs/5.3.473}
  {\path{doi:10.1093/rfs/5.3.473}}.
\newline\urlprefix\url{http://rfs.oxfordjournals.org/content/6/3/473.abstract}

\bibitem{Antweiler2004}
W.~Antweiler, M.~Z. Frank, \href{http://www.jstor.org/stable/info/3694736}{{Is
  All That Talk Just Noise? The Information Content of Internet Stock Message
  Boards}}, The Journal of Finance 59~(3) (2004) 1259--1294.
\newline\urlprefix\url{http://www.jstor.org/stable/info/3694736}

\bibitem{Sprenger2014}
T.~O. Sprenger, P.~G. Sandner, A.~Tumasjan, I.~M. Welpe,
  \href{http://doi.wiley.com/10.1111/jbfa.12086}{{News or Noise? Using Twitter
  to Identify and Understand Company-specific News Flow}}, Journal of Business
  Finance {\&} Accounting 41~(7-8) (2014) 791--830.
\newblock \href {http://dx.doi.org/10.1111/jbfa.12086}
  {\path{doi:10.1111/jbfa.12086}}.
\newline\urlprefix\url{http://doi.wiley.com/10.1111/jbfa.12086}

\bibitem{Vayanos2013}
D.~Vayanos, P.~Woolley,
  \href{https://academic.oup.com/rfs/article-lookup/doi/10.1093/rfs/hht014}{{An
  Institutional Theory of Momentum and Reversal}}, Review of Financial Studies
  26~(5) (2013) 1087--1145.
\newblock \href {http://dx.doi.org/10.1093/rfs/hht014}
  {\path{doi:10.1093/rfs/hht014}}.
\newline\urlprefix\url{https://academic.oup.com/rfs/article-lookup/doi/10.1093/rfs/hht014}

\bibitem{Hong1999}
H.~Hong, J.~C. Stein, \href{http://doi.wiley.com/10.1111/0022-1082.00184}{{A
  Unified Theory of Underreaction, Momentum Trading, and Overreaction in Asset
  Markets}}, The Journal of Finance 54~(6) (1999) 2143--2184.
\newblock \href {http://dx.doi.org/10.1111/0022-1082.00184}
  {\path{doi:10.1111/0022-1082.00184}}.
\newline\urlprefix\url{http://doi.wiley.com/10.1111/0022-1082.00184}

\bibitem{Chan2003}
W.~S. Chan,
  \href{http://www.sciencedirect.com/science/article/pii/S0304405X03001466}{{Stock
  price reaction to news and no-news: drift and reversal after headlines}},
  Journal of Financial Economics 70~(2) (2003) 223--260.
\newblock \href {http://dx.doi.org/10.1016/S0304-405X(03)00146-6}
  {\path{doi:10.1016/S0304-405X(03)00146-6}}.
\newline\urlprefix\url{http://www.sciencedirect.com/science/article/pii/S0304405X03001466}

\bibitem{Boudoukh2013}
J.~Boudoukh, R.~Feldman, S.~Kogan, M.~Richardson,
  \href{http://www.nber.org/papers/w18725}{{Which News Moves Stock Prices? A
  Textual Analysis}}, NBER Working Paper.
\newline\urlprefix\url{http://www.nber.org/papers/w18725}

\bibitem{Antoniou2013}
C.~Antoniou, J.~A. Doukas, A.~Subrahmanyam,
  \href{http://www.journals.cambridge.org/abstract{\_}S0022109012000592}{{Cognitive
  Dissonance, Sentiment, and Momentum}}, Journal of Financial and Quantitative
  Analysis 48~(01) (2013) 245--275.
\newblock \href {http://dx.doi.org/10.1017/S0022109012000592}
  {\path{doi:10.1017/S0022109012000592}}.
\newline\urlprefix\url{http://www.journals.cambridge.org/abstract{\_}S0022109012000592}

\bibitem{cci2011}
\href{https://www.conference-board.org/pdf{\_}free/press/TechnicalPDF{\_}4134{\_}1298367128.pdf}{{Consumer
  Confidence Survey -- technical note}}, Tech. rep. (2011).
\newline\urlprefix\url{https://www.conference-board.org/pdf{\_}free/press/TechnicalPDF{\_}4134{\_}1298367128.pdf}

\bibitem{Tetlock2007}
P.~C. Tetlock,
  \href{http://doi.wiley.com/10.1111/j.1540-6261.2007.01232.x}{{Giving Content
  to Investor Sentiment: The Role of Media in the Stock Market}}, The Journal
  of Finance 62~(3) (2007) 1139--1168.
\newblock \href {http://dx.doi.org/10.1111/j.1540-6261.2007.01232.x}
  {\path{doi:10.1111/j.1540-6261.2007.01232.x}}.
\newline\urlprefix\url{http://doi.wiley.com/10.1111/j.1540-6261.2007.01232.x}

\bibitem{Kogan2009}
S.~Kogan, D.~Levin, B.~R. Routledge, J.~S. Sagi, N.~A. Smith,
  \href{http://www.aclweb.org/anthology/N09-1031}{{Predicting Risk from
  Financial Reports with Regression}}, in: Annual Conference of the North
  American Chapter of the Association for Computational Linguistics, 2009, pp.
  272--280.
\newline\urlprefix\url{http://www.aclweb.org/anthology/N09-1031}

\bibitem{Wang2013}
C.-J. Wang, M.-F. Tsai, T.~Liu, C.-T. Chang,
  \href{http://www.aclweb.org/anthology/I13-1097}{{Financial Sentiment Analysis
  for Risk Prediction}}, in: International Joint Conference on Natural Language
  Processing, 2013, pp. 802--808.
\newline\urlprefix\url{http://www.aclweb.org/anthology/I13-1097}

\bibitem{Tsai2014}
M.-F. Tsai, C.-J. Wang, \href{http://aclweb.org/anthology/D14-1152}{{Financial
  Keyword Expansion via Continuous Word Vector Representations}}, in:
  Proceedings of the 2014 Conference on Empirical Methods in Natural Language
  Processing (EMNLP), Association for Computational Linguistics, Stroudsburg,
  PA, USA, 2014, pp. 1453--1458.
\newblock \href {http://dx.doi.org/10.3115/v1/D14-1152}
  {\path{doi:10.3115/v1/D14-1152}}.
\newline\urlprefix\url{http://aclweb.org/anthology/D14-1152}

\bibitem{Nopp2015}
C.~Nopp, T.~Wien, A.~Hanbury,
  \href{http://www.emnlp2015.org/proceedings/EMNLP/pdf/EMNLP071.pdf}{{Detecting
  Risks in the Banking System by Sentiment Analysis}}, in: Proceedings of the
  2015 Conference on Empirical Methods in Natural Language Processing,, 2015,
  pp. 591--600.
\newline\urlprefix\url{http://www.emnlp2015.org/proceedings/EMNLP/pdf/EMNLP071.pdf}

\bibitem{Rekabsaz2017}
N.~Rekabsaz, M.~Lupu, A.~Baklanov, A.~Hanbury, A.~Ur, L.~Anderson, T.~Wien,
  \href{https://doi.org/10.18653/v1/P17-1157}{{Volatility Prediction using
  Financial Disclosures Sentiments with Word Embedding-based IR Models}}, in:
  55th Annual Meeting of the Association for Computational Linguistics, 2017,
  pp. 1712--1721.
\newblock \href {http://dx.doi.org/10.18653/v1/P17-1157}
  {\path{doi:10.18653/v1/P17-1157}}.
\newline\urlprefix\url{https://doi.org/10.18653/v1/P17-1157}

\bibitem{Conneau2017}
A.~Conneau, D.~Kiela, H.~Schwenk, L.~Barrault, A.~Bordes,
  \href{http://arxiv.org/abs/1705.02364}{{Supervised Learning of Universal
  Sentence Representations from Natural Language Inference Data}}\href
  {http://arxiv.org/abs/1705.02364} {\path{arXiv:1705.02364}}, \href
  {http://dx.doi.org/10.1.1.156.2685} {\path{doi:10.1.1.156.2685}}.
\newline\urlprefix\url{http://arxiv.org/abs/1705.02364}

\bibitem{Mou2016}
L.~Mou, Z.~Meng, R.~Yan, G.~Li, Y.~Xu, L.~Zhang, Z.~Jin,
  \href{http://arxiv.org/abs/1603.06111}{{How Transferable are Neural Networks
  in NLP Applications?}}\href {http://arxiv.org/abs/1603.06111}
  {\path{arXiv:1603.06111}}.
\newline\urlprefix\url{http://arxiv.org/abs/1603.06111}

\bibitem{Howard2018}
J.~Howard, S.~Ruder, \href{http://arxiv.org/abs/1801.06146}{{Universal Language
  Model Fine-tuning for Text Classification}}\href
  {http://arxiv.org/abs/1801.06146} {\path{arXiv:1801.06146}}.
\newline\urlprefix\url{http://arxiv.org/abs/1801.06146}

\bibitem{Loughran2011}
T.~Loughran, B.~Mcdonald, \href{http://bit.ly/15GhT7K}{{When is a Liability not
  a Liability? Textual Analysis , Dictionaries , and 10-Ks}}, The Journal of
  Finance 66~(1) (2011) 35--65.
\newline\urlprefix\url{http://bit.ly/15GhT7K}

\bibitem{MikolovNIPS2013}
T.~Mikolov, I.~Sutskever, K.~Chen, G.~S. Corrado, J.~Dean,
  \href{https://dl.acm.org/citation.cfm?id=2999959}{{Distributed
  Representations of Words and Phrases and their Compositionality}}, in:
  C.~J.~C. Burges, L.~Bottou, M.~Welling, Z.~Ghahramani, K.~Q. Weinberger
  (Eds.), Advances in Neural Information Processing Systems 26, Curran
  Associates, Inc., 2013, pp. 3111--3119.
\newline\urlprefix\url{https://dl.acm.org/citation.cfm?id=2999959}

\bibitem{Baltrusaitis2017}
T.~Baltru{\v{s}}aitis, C.~Ahuja, L.-P. Morency,
  \href{http://arxiv.org/abs/1705.09406}{{Multimodal Machine Learning: A Survey
  and Taxonomy}}\href {http://arxiv.org/abs/1705.09406}
  {\path{arXiv:1705.09406}}.
\newline\urlprefix\url{http://arxiv.org/abs/1705.09406}

\bibitem{Bollen2011}
J.~Bollen, H.~Mao, X.-J. Zeng,
  \href{http://www.sciencedirect.com/science/article/pii/S187775031100007X}{{Twitter
  Mood Predicts the Stock Market}}, Journal of Computational Science 2~(1)
  (2011) 1--8.
\newblock \href {http://arxiv.org/abs/arXiv:1010.3003v1}
  {\path{arXiv:arXiv:1010.3003v1}}.
\newline\urlprefix\url{http://www.sciencedirect.com/science/article/pii/S187775031100007X}

\bibitem{Schumaker2009}
R.~P. Schumaker, H.~Chen,
  \href{http://doi.acm.org/10.1145/1462198.1462204}{{Textual Analysis of Stock
  Market Prediction Using Breaking Financial News: The AZFin Text System}}, ACM
  Trans. Inf. Syst. 27~(2) (2009) 12:1----12:19.
\newblock \href {http://dx.doi.org/10.1145/1462198.1462204}
  {\path{doi:10.1145/1462198.1462204}}.
\newline\urlprefix\url{http://doi.acm.org/10.1145/1462198.1462204}

\bibitem{Nguyen2015}
T.~H. Nguyen, K.~Shirai, \href{http://www.aclweb.org/anthology/P15-1131}{{Topic
  Modeling based Sentiment Analysis on Social Media for Stock Market
  Prediction}}, in: Proceedings of the 53rd Annual Meeting of the Association
  for Computational Linguistics, 2015, pp. 1354--1364.
\newline\urlprefix\url{http://www.aclweb.org/anthology/P15-1131}

\bibitem{Ding2015}
X.~Ding, Y.~Zhang, T.~Liu, J.~Duan,
  \href{https://www.ijcai.org/Proceedings/15/Papers/329.pdf}{{Deep learning for
  event-driven stock prediction}}, in: Proceedings of the 24th International
  Joint Conference on Artificial Intelligence (ICJAI 15), 2015, pp. 2327--2333.
\newline\urlprefix\url{https://www.ijcai.org/Proceedings/15/Papers/329.pdf}

\bibitem{Pinheiro2017}
L.~d.~S. Pinheiro, M.~Dras,
  \href{https://aclanthology.coli.uni-saarland.de/papers/U17-1001/u17-1001}{{Stock
  Market Prediction with Deep Learning: A Character-based Neural Language Model
  for Event-based Trading}}, in: Proceedings of the Australasian Language
  Technology Association Workshop 2017, 2017, pp. 6--15.
\newline\urlprefix\url{https://aclanthology.coli.uni-saarland.de/papers/U17-1001/u17-1001}

\bibitem{Deng2009}
J.~Deng, W.~Dong, R.~Socher, L.-J. Li, {Kai Li}, {Li Fei-Fei},
  \href{http://ieeexplore.ieee.org/document/5206848/}{{ImageNet: A large-scale
  hierarchical image database}}, in: 2009 IEEE Conference on Computer Vision
  and Pattern Recognition, IEEE, 2009, pp. 248--255.
\newblock \href {http://dx.doi.org/10.1109/CVPR.2009.5206848}
  {\path{doi:10.1109/CVPR.2009.5206848}}.
\newline\urlprefix\url{http://ieeexplore.ieee.org/document/5206848/}

\bibitem{Razavian2014}
A.~S. Razavian, H.~Azizpour, J.~Sullivan, S.~Carlsson,
  \href{http://ieeexplore.ieee.org/lpdocs/epic03/wrapper.htm?arnumber=6910029}{{CNN
  Features Off-the-Shelf: An Astounding Baseline for Recognition}}, in: 2014
  IEEE Conference on Computer Vision and Pattern Recognition Workshops, IEEE,
  2014, pp. 512--519.
\newblock \href {http://dx.doi.org/10.1109/CVPRW.2014.131}
  {\path{doi:10.1109/CVPRW.2014.131}}.
\newline\urlprefix\url{http://ieeexplore.ieee.org/lpdocs/epic03/wrapper.htm?arnumber=6910029}

\bibitem{Bowman2015}
S.~R. Bowman, G.~Angeli, C.~Potts, C.~D. Manning,
  \href{http://aclweb.org/anthology/D15-1075}{{A large annotated corpus for
  learning natural language inference}}, in: Proceedings of the 2015 Conference
  on Empirical Methods in Natural Language Processing, Association for
  Computational Linguistics, Stroudsburg, PA, USA, 2015, pp. 632--642.
\newblock \href {http://dx.doi.org/10.18653/v1/D15-1075}
  {\path{doi:10.18653/v1/D15-1075}}.
\newline\urlprefix\url{http://aclweb.org/anthology/D15-1075}

\bibitem{Lewis2004}
D.~D. Lewis, Y.~Yang, T.~G. Rose, F.~Li,
  \href{http://dl.acm.org/citation.cfm?id=1005332.1005345}{{RCV1: A New
  Benchmark Collection for Text Categorization Research}}, The Journal of
  Machine Learning Research 5 (2004) 361--397.
\newline\urlprefix\url{http://dl.acm.org/citation.cfm?id=1005332.1005345}

\bibitem{Engle1982}
R.~F. Engle,
  \href{https://www.jstor.org/stable/1912773?origin=crossref}{{Autoregressive
  Conditional Heteroscedasticity with Estimates of the Variance of United
  Kingdom Inflation}}, Econometrica 50~(4) (1982) 987.
\newblock \href {http://dx.doi.org/10.2307/1912773}
  {\path{doi:10.2307/1912773}}.
\newline\urlprefix\url{https://www.jstor.org/stable/1912773?origin=crossref}

\bibitem{Bollerslev1986}
T.~Bollerslev,
  \href{https://www.sciencedirect.com/science/article/pii/0304407686900631}{{Generalized
  autoregressive conditional heteroskedasticity}}, Journal of Econometrics
  31~(3) (1986) 307--327.
\newblock \href {http://dx.doi.org/10.1016/0304-4076(86)90063-1}
  {\path{doi:10.1016/0304-4076(86)90063-1}}.
\newline\urlprefix\url{https://www.sciencedirect.com/science/article/pii/0304407686900631}

\bibitem{Hansen2005}
P.~R. Hansen, A.~Lunde, \href{http://doi.wiley.com/10.1002/jae.800}{{A forecast
  comparison of volatility models: does anything beat a GARCH(1,1)?}}, Journal
  of Applied Econometrics 20~(7) (2005) 873--889.
\newblock \href {http://dx.doi.org/10.1002/jae.800}
  {\path{doi:10.1002/jae.800}}.
\newline\urlprefix\url{http://doi.wiley.com/10.1002/jae.800}

\bibitem{Andersen1998}
T.~G. Andersen, T.~Bollerslev,
  \href{https://www.jstor.org/stable/2527343?origin=crossref}{{Answering the
  Skeptics: Yes, Standard Volatility Models do Provide Accurate Forecasts}},
  International Economic Review 39~(4) (1998) 885.
\newblock \href {http://dx.doi.org/10.2307/2527343}
  {\path{doi:10.2307/2527343}}.
\newline\urlprefix\url{https://www.jstor.org/stable/2527343?origin=crossref}

\bibitem{Molnar2012}
P.~Molnar,
  \href{https://www.sciencedirect.com/science/article/pii/S1057521911000731}{{Properties
  of range-based volatility estimators}}, International Review of Financial
  Analysis 23 (2012) 20--29.
\newblock \href {http://dx.doi.org/10.1016/J.IRFA.2011.06.012}
  {\path{doi:10.1016/J.IRFA.2011.06.012}}.
\newline\urlprefix\url{https://www.sciencedirect.com/science/article/pii/S1057521911000731}

\bibitem{Pennington2014}
J.~Pennington, R.~Socher, C.~D. Manning,
  \href{https://nlp.stanford.edu/pubs/glove.pdf}{{GloVe: Global Vectors for
  Word Representation}}.
\newline\urlprefix\url{https://nlp.stanford.edu/pubs/glove.pdf}

\bibitem{Hochreiter1997}
S.~Hochreiter, J.~Schmidhuber,
  \href{http://www.mitpressjournals.org/doi/10.1162/neco.1997.9.8.1735}{{Long
  Short-Term Memory}}, Neural Computation 9~(8) (1997) 1735--1780.
\newblock \href {http://dx.doi.org/10.1162/neco.1997.9.8.1735}
  {\path{doi:10.1162/neco.1997.9.8.1735}}.
\newline\urlprefix\url{http://www.mitpressjournals.org/doi/10.1162/neco.1997.9.8.1735}

\bibitem{Gers2000}
F.~A. Gers, J.~Schmidhuber, F.~Cummins,
  \href{http://www.mitpressjournals.org/doi/10.1162/089976600300015015}{{Learning
  to Forget: Continual Prediction with LSTM}}, Neural Computation 12~(10)
  (2000) 2451--2471.
\newblock \href {http://dx.doi.org/10.1162/089976600300015015}
  {\path{doi:10.1162/089976600300015015}}.
\newline\urlprefix\url{http://www.mitpressjournals.org/doi/10.1162/089976600300015015}

\bibitem{Schuster1997}
M.~Schuster, K.~Paliwal,
  \href{http://ieeexplore.ieee.org/document/650093/}{{Bidirectional recurrent
  neural networks}}, IEEE Transactions on Signal Processing 45~(11) (1997)
  2673--2681.
\newblock \href {http://dx.doi.org/10.1109/78.650093}
  {\path{doi:10.1109/78.650093}}.
\newline\urlprefix\url{http://ieeexplore.ieee.org/document/650093/}

\bibitem{Lai2015}
S.~Lai, L.~Xu, K.~Liu, J.~Z. AAAI, U.~2015,
  \href{http://www.aaai.org/ocs/index.php/AAAI/AAAI15/paper/download/9745/9552}{{Recurrent
  Convolutional Neural Networks for Text Classification.}}, in: AAAI, 2015, pp.
  2267--2273.
\newline\urlprefix\url{http://www.aaai.org/ocs/index.php/AAAI/AAAI15/paper/download/9745/9552}

\bibitem{Li2016}
P.~Li, W.~Li, Z.~He, X.~Wang, Y.~Cao, J.~Zhou, W.~Xu,
  \href{https://arxiv.org/pdf/1607.06275.pdf}{{Dataset and Neural Recurrent
  Sequence Labeling Model for Open-Domain Factoid Question Answering}}.
\newline\urlprefix\url{https://arxiv.org/pdf/1607.06275.pdf}

\bibitem{Liu2016}
Y.~Liu, C.~Sun, L.~Lin, X.~Wang,
  \href{https://arxiv.org/pdf/1605.09090.pdf}{{Learning Natural Language
  Inference using Bidirectional LSTM model and Inner-Attention}}\href
  {http://arxiv.org/abs/arXiv:1605.09090v1} {\path{arXiv:arXiv:1605.09090v1}}.
\newline\urlprefix\url{https://arxiv.org/pdf/1605.09090.pdf}

\bibitem{Lin2017}
Z.~Lin, M.~Feng, C.~Nogueira, D.~Santos, M.~Yu, B.~Xiang, B.~Zhou, Y.~Bengio,
  \href{https://arxiv.org/pdf/1703.03130.pdf}{{A Structured Self-Attentive
  Sentence Embedding}}, in: ICLR, 2017.
\newline\urlprefix\url{https://arxiv.org/pdf/1703.03130.pdf}

\bibitem{Lipton2016}
Z.~C. Lipton, D.~C. Kale, C.~Elkan, R.~Wetzel,
  \href{http://arxiv.org/abs/1511.03677}{{Learning to Diagnose with LSTM
  Recurrent Neural Networks}}, in: ICLR, 2016.
\newblock \href {http://arxiv.org/abs/1511.03677} {\path{arXiv:1511.03677}}.
\newline\urlprefix\url{http://arxiv.org/abs/1511.03677}

\bibitem{LiptonClinical2016}
Z.~C. Lipton, D.~Kale, R.~Wetzel,
  \href{http://proceedings.mlr.press/v56/Lipton16.html}{{Directly Modeling
  Missing Data in Sequences with RNNs: Improved Classification of Clinical Time
  Series}}, in: Proceedings of the 1st Machine Learning for Healthcare
  Conferenc, 2016, pp. 253--270.
\newline\urlprefix\url{http://proceedings.mlr.press/v56/Lipton16.html}

\bibitem{Alberg2017a}
J.~Alberg, Z.~C. Lipton, \href{http://arxiv.org/abs/1711.04837}{{Improving
  Factor-Based Quantitative Investing by Forecasting Company Fundamentals}},
  in: 31st Conference on Neural Information Processing Systems (NIPS), 2017.
\newblock \href {http://arxiv.org/abs/1711.04837} {\path{arXiv:1711.04837}}.
\newline\urlprefix\url{http://arxiv.org/abs/1711.04837}

\bibitem{Kingma2015}
D.~P. Kingma, J.~{Lei Ba}, \href{https://arxiv.org/pdf/1412.6980.pdf}{{Adam: A
  method for stochastic optimization}}, in: ICLR, 2015.
\newblock \href {http://arxiv.org/abs/arXiv:1412.6980v9}
  {\path{arXiv:arXiv:1412.6980v9}}.
\newline\urlprefix\url{https://arxiv.org/pdf/1412.6980.pdf}

\bibitem{Johnson2015}
R.~Johnson, T.~Zhang,
  \href{http://www.anthology.aclweb.org/N/N15/N15-1011.pdf}{{Effective Use of
  Word Order for Text Categorization with Convolutional Neural Networks}}, in:
  NAACL, 2015, pp. 103--112.
\newline\urlprefix\url{http://www.anthology.aclweb.org/N/N15/N15-1011.pdf}

\end{thebibliography}

\end{document}